\documentclass[journal=apchd5]{achemso}
\usepackage[version=3]{mhchem} 
\usepackage[T1]{fontenc}       

\usepackage{graphicx}
\usepackage{dcolumn}
\usepackage{bm}
\usepackage{mathtools}
\usepackage{hyperref}
\usepackage{color}
\usepackage[a4paper]{anysize}

\newcommand{\A}{{\bf A}}

\renewcommand{\j}{{\bf j}}

\newcommand{\I}{{\rm i}}
 
\newcommand{\re}{\ensuremath{\mathrm{Re}\,}}
\newcommand{\im}{\ensuremath{\mathrm{Im}\,}}

\def\gsim{\lower.35em\hbox{$\stackrel{\textstyle>}{\textstyle\sim}$}}
\def\lsim{\lower.35em\hbox{$\stackrel{\textstyle<}{\textstyle\sim}$}}

\title{Plasmonics in topological insulators: Spin-charge separation, the influence of the inversion layer, and phonon-plasmon coupling} 
\author{Tobias Stauber} 
\email{tobias.stauber@csic.es}
\affiliation{Departamento de Teor\'{\i}a y Simulaci\'on de Materiales, Instituto de Ciencia de Materiales de Madrid, CSIC, 28049 Cantoblanco, Spain}
\author{Guillermo G\'omez-Santos}
\affiliation{Departamento de F\'{\i}sica de la Materia Condensada, INC and IFIMAC, Universidad Aut\'onoma de Madrid, E-28049 Madrid, Spain}
\author{Luis Brey}
\affiliation{Departamento de Teor\'{\i}a y Simulaci\'on de Materiales, Instituto de Ciencia de Materiales de Madrid, CSIC, 28049 Cantoblanco, Spain}
\begin{document}
\begin{abstract}
We demonstrate via three examples that topological insulators (TI) offer a new platform for plasmonics. First, we show that the collective excitations of a thin slab of a TI display spin-charge separation. This gives rise to purely charge-like optical and purely spin-like acoustic plasmons, respectively. Second, we argue that the depletion layer mixes Dirac and Schr\"odinger electrons which can lead to novel features such as high modulation depths and interband plasmons. The analysis is based on an extension of the usual formula for optical plasmons that depends on the slab width and on the dielectric constant of the TI. Third, we discuss the coupling of the TI surface phonons to  the plasmons and find strong hybridisation especially for samples with large slab widths.
\end{abstract}
\maketitle
\section{Introduction}
Three-dimensional topological insulator (TI) like Bi${}_2$Se${}_3$ or Bi${}_2$Te${}_3$ present a new class of plasmonic materials that have the potential to add new aspects to the already well-established systems including noble metals or two-dimensional materials.\cite{Basov16,Low17} These properties can be linked to the atomistic structure of the TIs that are built up from hexagonal unit cells containing 5 two-dimensional sheets.\cite{Hsieh09,Chen09,Xia09} Further, they display protected surface states due to the strong spin-orbit coupling which will give rise to spin-charge separation of collective excitations.\cite{Stauber13}

One of the main differences between Dirac carriers at the surface of a TI and Dirac carrier in graphene is that in the case of graphene momentum and pseudo-spin are correlated giving rise to phenomenons such as Klein tunneling. But in the case of TIs, it is momentum and real spin which are locked.\cite{Hasan10,Qi11} This difference will give rise to interesting new phenomena not found in graphene-based systems as first pointed out by Zhang and co-workers.\cite{Raghu10} They discussed the collective modes of this "helical metal" focusing on the curious fact that density fluctuations induce transverse spin fluctuations and vice versa. A transverse spin wave can be generated by a transient spin grating consisting of two orthogonally polarized non collinear incident beams.\cite{Raghu10} To detect the induced charge density wave, one measures e.g. the spatial modulation of reflectivity. These spin-plasmons were also discussed in terms of the plasmon wave function.\cite{Efimkin12,Efimkin12b}  In this article, we will review recent results on plasmonic excitations in topological insulators and show that there are small, but far-reaching differences. We also include a discussion on the influence of the TI phonon modes on the plasmon dispersion. 

Surface plasmons (or plasmon-polaritons) are collective charge oscillations which are usually confined to the surface of a three-dimensional (3D) metallic material which has an interface with an insulator, typically air. Their maximal energy $\hbar\omega_p$ is related to the bulk plasmons of energy $\omega_p^{3D}=\frac{e^2n}{\epsilon_0m}$ and given by $\omega_p=\omega_p^{3D}/\sqrt{2}$ with particle density $n$ and electron mass $m$.\cite{Ritchie57} In the case of topological insulators, we encounter a slightly different situation, i.e., there is a 2D metallic surface consisting of the topologically protected surface states (TSS) and/or a 2D depletion layer surrounded by two insulators: one is given by the topological bulk insulator and the other one is usually air. The plasmonic dispersion is thus characterised by the typical square-root behavior of 2D metals:\cite{Stauber14}
\begin{align}
\omega_p=\sqrt{\frac{D}{2\epsilon_0\epsilon}q}\;,
\label{Plasmon2D}
\end{align}
where $D$ is the Drude weight or charge stiffness and $\epsilon$ the effective dielectric function. Further, $\epsilon_0$ is the vacuum permittivity and $q$ the two-dimensional wave number. 

The plasmonic dispersion for small $q$ is thus characterised by the Drude weight $D$ and the dielectric medium $\epsilon$. For a 2D electron gas (2DEG), we have $D_{2DEG}=e^2n/m$. For Dirac fermions with spin and valley-degeneracy $g_s$ and $g_v$, we have $D_{Dirac}=\frac{e^2v_F}{\hbar}\sqrt{\frac{g_sg_vn}{4\pi}}$, i.e., for a topological insulator with TSS, $D_{Top}=\frac{e^2v_F}{\hbar}\sqrt{\frac{n}{4\pi}}$. The effective dielectric function is often given by $\epsilon=(\epsilon_T+\epsilon_B)/2$, with the top dielectric layer usually assumed to be vacuum ($\epsilon_T=1$) and the bottom dielectric layer $\epsilon_B>1$. 

There have been a number of experimental works which investigated the optical response and plasmonic resonances of bulk topological insulators\cite{Valdes12,DiPietro13,Autore15a,Autore15b,Post15,Sim15,Zhao15,Politano15,Wu15,Autore16,Viti16,Autore17,Politano17} and new theoretical phenomena were predicted.\cite{Juergens14a,Juergens14b,Qi15,Wu15b} Also, plasmons in topological nano particles have been dealt with.\cite{Lin15,Siroki17} Here, we will focus on two aspects which we have discussed previously in Ref. \cite{Stauber13,Stauber14}. They are related to a thin slab of a topological insulator and involve the top and bottom surface that are separated by a strong dielectric defined by the width of the TI. These structures can be ideally modelled as a double layer of Dirac electrons and we showed that there is spin-charge separation of the collective excitation in the case of equal electronic density of the upper and lower layer. We also pointed out that the plasmonic modes usually depend on the sample thickness and on the dielectric constant of the bulk - only for very small wave numbers, the universal Dirac dispersion is recovered. In order to explain recent experiments, we proposed to also consider the influence of the depletion layer underneath the two Dirac systems which is Schr\"odinger-like.

The review is organized as follows. First, we introduce the macroscopic as well as microscopic description of a thin slab of topological insulators. We then present analytical results exemplifying the spin-charge separation of the in-phase and out-of-phase mode. We also calculate the plasmon dispersion numerically and analyze the spin-charge separation for general wave numbers including a discussion for possible experimental set-ups to detect them. In Sec. V,  we critically compare our results with experiment and in Sec. VI we finally consider the influence of the TI's phononic mode on the plasmonic resonances. The appendix gives a general derivation of the optical f-sum rule needed for the experimental analysis of the plasmons. 
\section*{Response of topological insulators}
\subsection*{Macroscopic description}
\label{Macroscopic}
The electromagnetic response of typical topological insulators such as Bi$_2$Se$_3$, Bi$_2$Te$_3$ and Sb$_2$Te$_3$ in the far and mid-IR can be well described by an isotropic Drude-Lorentz model,
\begin{align}
\label{Drude-Lorentz}
\epsilon(\omega)=\epsilon_\infty+\sum_{j=1}^N\frac{\omega_{p,j}^2}{\omega_{0,j}^2-\omega^2-i\omega\gamma_j}\;.
\end{align}
Above, we assumed $N$ effective oscillators with $\omega_{p,j}$ being related to the oscillator strength or Drude weight of the $j$-th mode and $\omega_{0,j}$ being the resonance frequency with damping term $\gamma_j$. Finally, $\epsilon_\infty$ denotes the high-frequency dielectric constant which could also be modelled as one of the effective oscillators.

\begin{figure}
\centering
  \includegraphics[width=0.9\columnwidth]{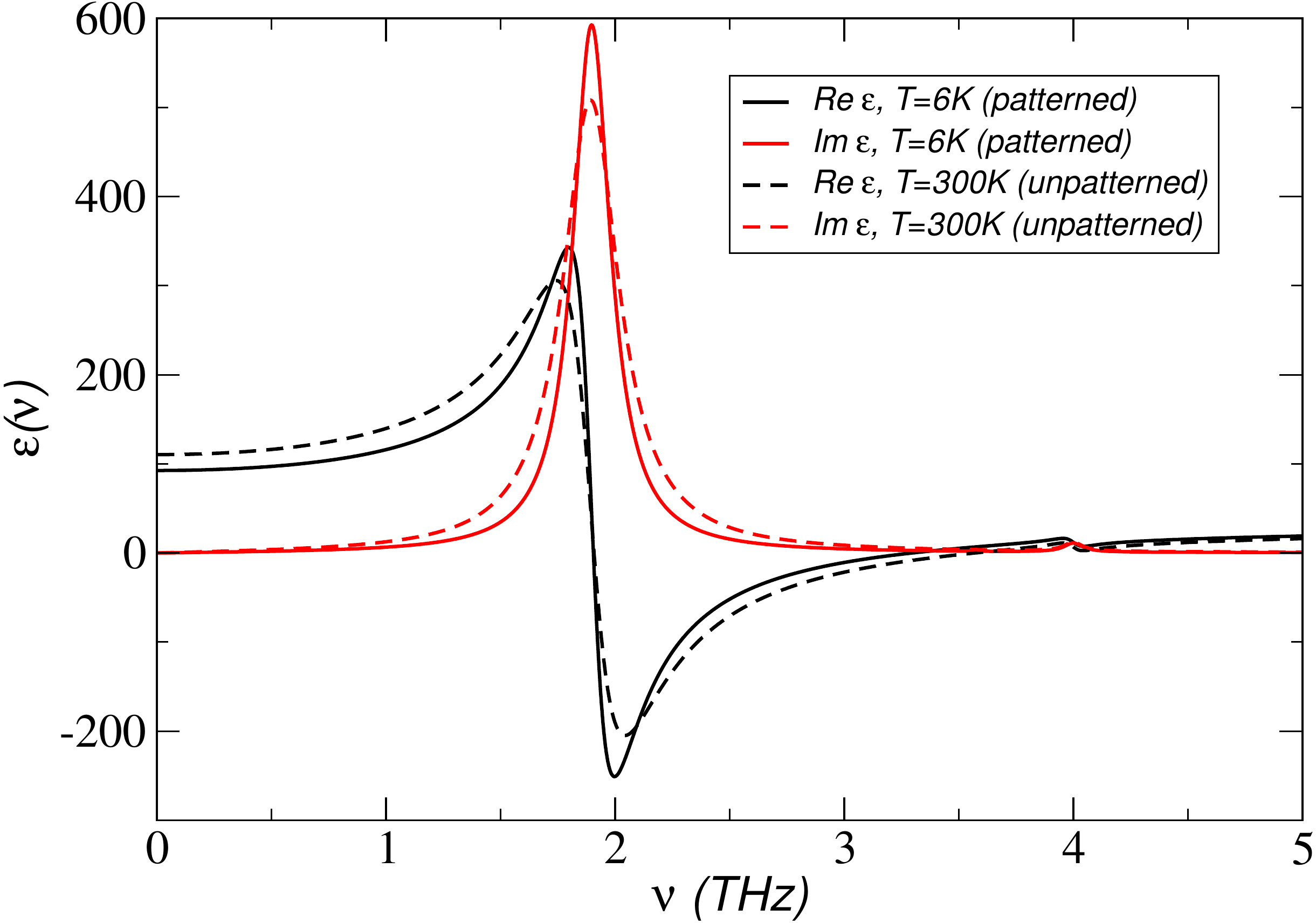}
\caption{The real (black) and imaginary (red) part of the dielectric function for $T=6$K (solid curves) and $T=300$K (dashed curves). Parameters taken from Ref. \cite{DiPietro13} with $\epsilon_\infty=30$.}
  \label{Dielectric}
\end{figure}

In the case of Bi$_2$Se$_3$, there are two maxima at 1.85 THz and 4.0 THz corresponding to the $\alpha$ and $\beta$ bulk phonon-modes, respectively. In Fig. \ref{Dielectric}, we show the real and imaginary part of the dielectric function as black and red curves, respectively. The solid (dashed) curves correspond to (un)patterned samples at $T=6$K ($T=300$K) and the parameters were taken from Ref. \cite{DiPietro13}. Models only including the $\alpha$-phonon also yield good fits to the experimental data.\cite{Valdes12} In the last section, we will use this model in order to also discuss the coupling of the surface phonons to the plasmons. 

Within the random-phase approximation, the relation to the conductivity is given by 
\begin{align}
\epsilon(\omega)=1+\frac{\I\sigma(\omega)}{\epsilon_0\epsilon\omega}\;.
\end{align}
The two-dimensional conductivity of the surface states is now obtained by superimposing to the model of Eq. (\ref{Drude-Lorentz}) a Drude-like component which is associated to the contribution of the topologically protected surface states as well as possible 2DEG-Fermions due to a depletion layer.\cite{Valdes12} Dividing the Drude-like component by the sample width and with the help of the f-sum rule, the corresponding electronic density as well as the effective mass is obtained. For details on the f-sum rule, see the appendix.

\subsection*{Microscopic description}
For a microscopic description, we model the topological insulator by two 2D Dirac systems separated by a dielectric spacer. The Hamiltonian of the top (T) and bottom (B) is given by 
\begin{equation}
\label{Hamiltonian}
H^{T/B} = \pm\hbar v_F  \left( \begin{array}{cc}
0 & -k_x -  i k_y  \\
-k_x +i k_y  & 0  
\end{array} \right) \;.
\end{equation} 
Important for the following is the different sign of the Fermi velocity $v_F$ between the top and the bottom Hamiltonians discussed in Refs. \cite{Lee09,Silvestrov12}.

\subsubsection*{Charge response function}
The charge response shall be modelled by a 4x4 matrix consisting of the response of the top and the bottom surface:
\begin{equation}
\chi _0 (q,\omega) =  \left( \begin{array}{cc}
\chi _0 ^T  & 0  \\
0  & \chi _0^B  
\end{array} \right)\;.
\end{equation}
The response functions $\chi_0^{T/B}$ are themselves matrices representing the charge and transverse spin degree of freedom. In topological insulators, the response of the charge density, $\rho$, and the response of the transverse spin density, $S_{\perp}$ are related and we can write\cite{Raghu10,Efimkin12,Efimkin12b}
\begin{equation}
\chi _0 ^{T/B} (q,\omega)  =   \left( \begin{array}{cc}
1 & \pm x  \\
\mp x  & -x^2  
\end{array} \right) \chi _{\rho\rho}^{T/B} (q,\omega)
\label{chi0}
\end{equation}
where $x= \frac  {\omega}{q v_F}$ and $\chi _{\rho\rho}^{T/B} (q,\omega)$ the (scalar) density-density correlation function of a single Dirac cone corresponding to the top/bottom surface. Note that there are also non-diagonal terms due to the topological spin-charge coupling. These terms change sign which can be traced back to the different sign of the Fermi velocity on the two surfaces, see Eq. (\ref{Hamiltonian}).

Plasmons always lie above the single particle dispersion, i.e., $\omega>v_Fq$. For clean systems ($\im\chi _{\rho\rho}^{T/B}=0$), there is an analytical formula for $\chi _{\rho\rho}^{T/B}$:\cite{Wunsch06,Hwang07} 
\begin{eqnarray}
\label{FullResponse}
\chi_{\rho\rho}^{T/B}&=&-\frac{\mu_{T/B}}{2\pi (\hbar v_F)^2}+\frac{1}{16\hbar\pi}\frac{q^2}{\sqrt{\omega^2-(v_Fq)^2}}\\\nonumber
&\times&\left[G\left(\frac{2\mu_{T/B}+\hbar\omega}{\hbar v_Fq}\right)-G\left(\frac{2\mu_{T/B}-\hbar\omega}{\hbar v_Fq}\right)\right]\;,
\end{eqnarray}
with $G(x)=x\sqrt{x^2-1}-\cosh^{-1}(x)$ for $x>1$ and $\mu_{T/B}$ are the chemical potentials of the top/bottom surface, respectively. 

\subsubsection*{Random-phase approximation and plasmons}
To discuss plasmons, we apply the random-phase approximation (RPA) and only the charge sectors of the opposite surfaces will be coupled. The system thus decouples into charge and transverse spin sector, but in order to make the spin-charge separation more explicit, we will keep the full 4x4 tensorial structure. The susceptibility in RPA then reads
\begin{equation}
\label{ResponseFunction}
\chi^{RPA}= \chi _0 (q,\omega)\left [ 1-v(q)\chi _0 (q,\omega) \right ] ^{-1}
\end{equation}
where
\begin{equation}
v(q)=\left(\begin{array}{cccc}
v_T(q) & 0 & v_{TB}(q)& 0  \\
0  & 0 & 0& 0 \\
v_{TB}(q)& 0 & v_B(q)& 0 \\
0&0&0&  0
\end{array}    \right )\;.
\end{equation}
Above, we defined $v_{T/B}(q)$ and $v_{TB}(q)$ as the intra- and interlayer Coulomb interaction of the top and bottom, respectively.  

In the following, we will focus on the general case with different dielectric media on the top ($\epsilon_T$), center ($\epsilon_{TI}$) and bottom ($\epsilon_B$). The expressions for the intra- and interlayer Coulomb potential are given by\cite{Profumo12,Badalyan12,Scharf12,Stauber12} $v_{T/B}=[\cosh(qd)+(\epsilon_{B/T}/\epsilon_{TI})\sinh(qd)]v_{TB}(q)$ and $v_{TB}=e^2\epsilon_{TI}/(\varepsilon_0qN)$  with $N=\epsilon_{TI}(\epsilon_T+\epsilon_B)\cosh(qd)+(\epsilon_T\epsilon_B+\epsilon_{TI}^2)\sinh(qd)$. The two surfaces are separated by the distance $d$ and there will always be a hybridization between the two plasmon modes for small enough $q$.\footnote{The general formulas {\it including retardation} effects are given in Refs. \cite{Stauber12, Stauber14}.}

To discuss the hybridisation of the plasmonic modes of the two layers, we need to calculate the zeros of $\det(1-v(q)\chi _0)$,
\begin{equation}
\label{det}
(1-v_T\chi_{\rho\rho} ^T)( 1-v_B\chi_{\rho\rho} ^B) - v _{TB} ^2 \chi_{\rho\rho} ^T \chi_{\rho\rho} ^B=0\;.
\end{equation}
In the following, we will solve Eq. (\ref{FullResponse}) analytically as well as numerically. 

\section*{Acoustic and optical modes}
The electrostatic interaction will couple the top and the bottom plasmonic excitations to form bonding and anti-bonding modes. These modes can be discussed analytically, using the long-wavelength ($q\to0$) or local approximation of the charge response
\begin{equation}
\label{local}
\chi_{\rho\rho}^{T/B}=\frac{\mu_{T/B}}{4\pi\hbar^2}\frac{q^2}{\omega^2}\;,
\end{equation}
valid for $\omega\gg v_Fq$ and $\mu_{T/B}\gsim\hbar\omega$. For the small parameter $qd\ll1$, we obtain analytical expressions for the optical (in-phase, "+") and acoustic (out-of-phase, "-") mode 
\begin{eqnarray}
\label{OpticalMode}
\omega _ + ^2 &  = & \frac {\alpha_d v_F(\mu _T + \mu _B )}{\hbar(\epsilon_T+\epsilon_B)} q\;, \\\label{acousticMode}
\omega _- ^2 &  = & \frac {\alpha_d v_F}{ \hbar\epsilon_{TI}} \frac {\mu _T  \mu _B } {\mu _T+ \mu _B } dq^2 \equiv v_s^2q^2\;.
\end{eqnarray}
Above, we defined the fine-structure of a general Dirac system $\alpha_d=\frac{e^2}{4\pi\varepsilon_0\hbar v_F}$ and introduced the sound velocity, $v_s$, in case of the acoustic (out-of-phase) mode.

\subsection*{General formula for acoustic mode}
The expression for the out-of-phase mode, Eq. (\ref{acousticMode}), has limitations and can only be applied if $(k_F^T+k_F^B)d/\epsilon_{TI}\gg1$, where $k_F^{T/B}=\mu_{T/B}/v_F$ denotes the Fermi wave number of the top and bottom surfaces. The general case involves the Laurent-Taylor expansion\cite{Santoro88} and the square root singularity of $\chi_{\rho\rho}$ at $\omega=v_Fq$ then guarantees that $v_s>v_F$.\cite{Stauber12} For a system with equal density at the top and at the bottom, $k_F^T=k_F^B\equiv k_F$, this reads
\begin{equation}
\label{AcousticModeTwo}
v_s=\frac{1+\alpha_dk_Fd/\epsilon_{TI}}{\sqrt{1+2\alpha_dk_Fd/\epsilon_{TI}}}v_F\;.
\end{equation}
For the general case, see Ref. \cite{Profumo12}.

There are regimes where the sound velocity is well inside the Pauli-blocked (protected) region. This mode can then be excited by a dipole close to the slab since it will predominately couple to large momenta.\cite{Ameen17}

\subsection*{General formula for optical mode}
Also the analytical formula for the optical mode, Eq. (\ref{OpticalMode}), cannot be applied without the necessary caution. Its applicability depends on the relative value of $\epsilon_T,\epsilon_B$ with respect to $\epsilon_{TI}$: only if they are of the same order of magnitude, it holds. The general formula valid for $qd\ll1$ is given by\cite{Stauber14}
\begin{equation}
	\frac{\omega_+^2}{g_sg_v\alpha_dv_F^2q}=\frac{\epsilon_2(k_F^T+k_F^B)+qd(\epsilon_Tk_F^B+\epsilon_Bk_F^T)+\sqrt{R}}{2\left[\epsilon_{TI}(\epsilon_T+\epsilon_B)+qd(\epsilon_{TI}^2+\epsilon_T\epsilon_B)\right]}
	\label{GeneralFormula}
\end{equation}
with 
\begin{align}
R=\epsilon_{TI}^2(k_F^T+k_F^B)^2-2qd\epsilon_{TI}(k_F^T-k_F^B)(\epsilon_Tk_F^B-\epsilon_Bk_F^T)+(qd)^2(\epsilon_Tk_F^B-\epsilon_Bk_F^T)^2\;.
\end{align} 
For a 3D topological insulator with $\epsilon_{TI}\gg\epsilon_T,\epsilon_B$ and in the case of equal densities $k_F^T=k_F^B=k_F$, this simplifies to 
\begin{equation}
\label{OpticalModeTwo}
\omega_+^2=\frac {2g_sg_v\alpha_d v_F^2k_Fq}{\epsilon_T+\epsilon_B}\left[1+\frac{qd\epsilon_{TI}}{\epsilon_T+\epsilon_B}\right]^{-1}\;.
\end{equation}
In this case, Eq. (\ref{OpticalMode}) is only valid for $qd\epsilon_{TI}/(\epsilon_T+\epsilon_B)\ll1$.  Let us again emphasise that Eqs. (\ref{GeneralFormula}) and (\ref{OpticalModeTwo}) are the formulas that must be used when discussing the plasmonic excitations of general topological insulators since $\epsilon_{TI}\gg1$ for general frequencies.

The above dependence of the plasmonic mode on the slab-width $d$ and on the relative ratio $\epsilon_{TI}/(\epsilon_T+\epsilon_B)$ was recently obtained for a thin slab using the Keldysh potential.\cite{Bondarev17} This exemplifies the general nature of this formula and we expect it to be also present in other quasi-2D materials with large internal dielectric medium.
\section*{Charge and Spin separation}
Plasmonic excitations in topological insulators are similar to the ones in double layer graphene because the relevant charge response is identical except from the degeneracy factor $g_sg_v=4$ consisting of the spin and valley degeneracy, only present in graphene. But there is another interesting signature only present in the helical metal which is related the spin-momentum locking. 

The key insight is that the Dirac cone on one TI surface is {\it not} an identical copy of the Dirac cone on the other surface because the sign of the Fermi velocity must be opposite for the two Dirac cones.\cite{Silvestrov12} This can be seen by mapping the two-dimensional slab onto the topologically equivalent finite sphere.\cite{Lee09} This mapping locks the orientation of the Pauli-spin matrices to be perpendicular with respect to the local surface and when an electron moves in a closed loop on the surface of an isotropic spherical topological insulator, the wave function acquires a Berry phase which is equal to the magnetic flux enclosed by an effective magnetic monopole of strength 1/2 located in the center of the sphere. Therefore, the rotation of the Pauli matrices induces a change of sign in the Dirac Hamiltonian when the path encloses a solid angle $2\pi$, i.e., going from the top (north pole) to the bottom (south pole) surface.

\begin{figure}
\centering
  \includegraphics[width=0.9\columnwidth]{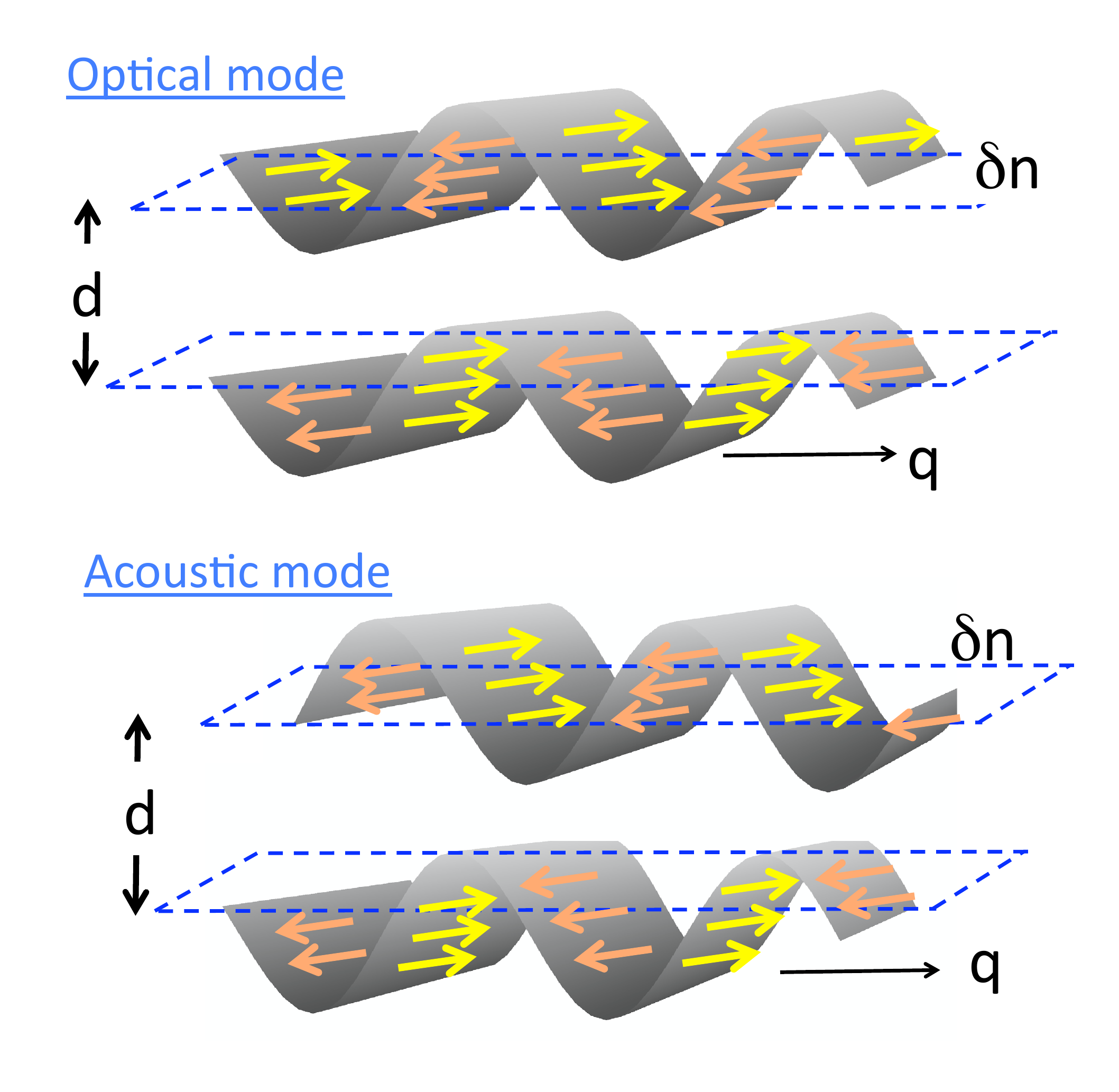}
  \caption{Schematic view of the spin-charge separation (Figure adapted from Ref. \cite{Stauber13}). The optical mode shown in the upper panel consists of in-phase charge oscillations and out-of-phase transverse spin oscillations with respect to the top and bottom layer. This yields an effective (pure) charge wave. The acoustic mode shown in the lower panel consists of out-of-phase charge oscillations and in-phase spin oscillations. This yields an effective (pure) spin wave.}
  \label{SpinChargeSep}
\end{figure}

The spin locked to the charge momentum is therefore polarized in contrary directions on the two sides. This has the consequence that the optical (in-phase) and acoustic (out-of-phase) oscillations can be purely charge- and spin-like, respectively, see Fig. \ref{SpinChargeSep}. The optical plasmon excitations detected using infrared spectroscopy\cite{DiPietro13} have thus a purely charge-like character without the spin-component. This might be the reason why the transverse spin component of the plasmons predicted for single layer\cite{Raghu10} has not been detected yet.  

\subsection*{Analytical results}
Let us discuss the spin-charge separation analytically. For plasmons, which are self-sustained excitation, the external potential is set to zero and the induced potential, therefore, reads $V^{ind} =v(q)\chi _0 (q,\omega) V^{ind}$. The collectives modes are given by the zeros of $\det(1-v(q)\chi _0)$ and the induced potential is thus obtained from 
\begin{equation}
(1-v(q)\chi _0)_{\omega =\omega_{\pm}} V^{ind}=0\;.
\end{equation}
From this, we obtain $V^{ind}$ for the bonding (+) and anti-bonding (-) modes:
\begin{equation}
V_+ ^{ind} = V_0\left (   \begin{array}{c}
1\\
0 \\
1\\
 0
\end{array}    \right  )\, \, \, {\rm and} \, \, \,
V _-^{ind} = V_0\left (   \begin{array}{c}
1\\
0 \\
-\frac {\mu _T} {\mu _B}\\
 0
\end{array}    \right  )
\end{equation}
From the induced potentials, we calculate the charge and spin densities via $\rho= \chi _0 (q,\omega)  V^{ind}$. For the optical mode, $\omega _+ $, this yields in the limit $q\to0$
\begin{equation}
\label{OpticalCharge}
\left (   \begin{array}{c}
\rho ^T  \\
S _{\perp} ^T \\
\rho ^B\\
  S _{\perp} ^B
\end{array}    \right  ) =\rho_T\left (   \begin{array}{c}
  1\\
-\sqrt { \frac {\alpha_d (\mu _T+\mu _B)  } {\hbar v_F q(\epsilon_T+\epsilon_B)}}\\
\frac{\mu_B}{\mu_T}\\
 \sqrt { \frac {\alpha_d (\mu _T+\mu _B)  } {\hbar v_F q(\epsilon_T+\epsilon_B)}}\frac{\mu_B}{\mu_T}
 \end{array}    \right  )\;.
\end{equation}
For the acoustic mode, $\omega _-$, one obtains
\begin{equation}
\label{AcousticSpin}
\left (   \begin{array}{c}
\rho ^T  \\
S _{\perp} ^T \\
\rho ^B\\
  S _{\perp} ^B
\end{array}    \right  ) =\rho_T\left (   \begin{array}{c}
 1\\
- \sqrt {\frac {\alpha_d d}{\hbar v_F\epsilon_{TI}} \frac {\mu _T \mu _B  } {\mu _T+ \mu _B }}\\
-1\\
- \sqrt {\frac {\alpha_d d}{\hbar v_F\epsilon_{TI}} \frac {\mu _T \mu _B  } {\mu _T+ \mu _B }}
 \end{array}    \right  )\;.
\end{equation}

Usually, the exciting light has a wavelength $\lambda$ which is larger than the slab thickness $d$. The composite charge and spin densities are given by the sum of the two surfaces, $\rho_c=\rho^T+\rho^B$ and $\rho_s=S_\perp^T+S_\perp^B$. The collective excitations of a thin TI slab with the same chemical potential on the top and on the bottom are purely charge- or spin-like, see Fig. \ref{SpinChargeSep}. 
Only for larger TI slabs, the optical and acoustic mode merge and the single-layer plasmon carrying charge and spin is recovered. 

We finally note that the  optical mode looses its pure charge character if different charge densities $\mu_T\neq\mu_B$ are present. Still, for the acoustic mode, the pure spin character is preserved.

\subsection*{Numerical results}
To see if spin-charge separation is stable at larger wave numbers, we will solve the modes defined by Eq. (\ref{det}) and Eq. (\ref{FullResponse}) numerically. We choose parameters corresponding to $v_F=6\times10^5$m/s and $n_{T/B}=1.5\times10^{13}$cm${}^{-2}$.\cite{Bansal12,DiPietro13} We also set $\epsilon_B=10$, $\epsilon_T=1$ and $\epsilon_{TI}=100$ as done in Ref. \cite{Profumo12}, even though a smaller value $\epsilon_{TI}=30$ could also be used.\cite{Autore17} 

\begin{figure}
\centering
  \includegraphics[width=0.99\columnwidth]{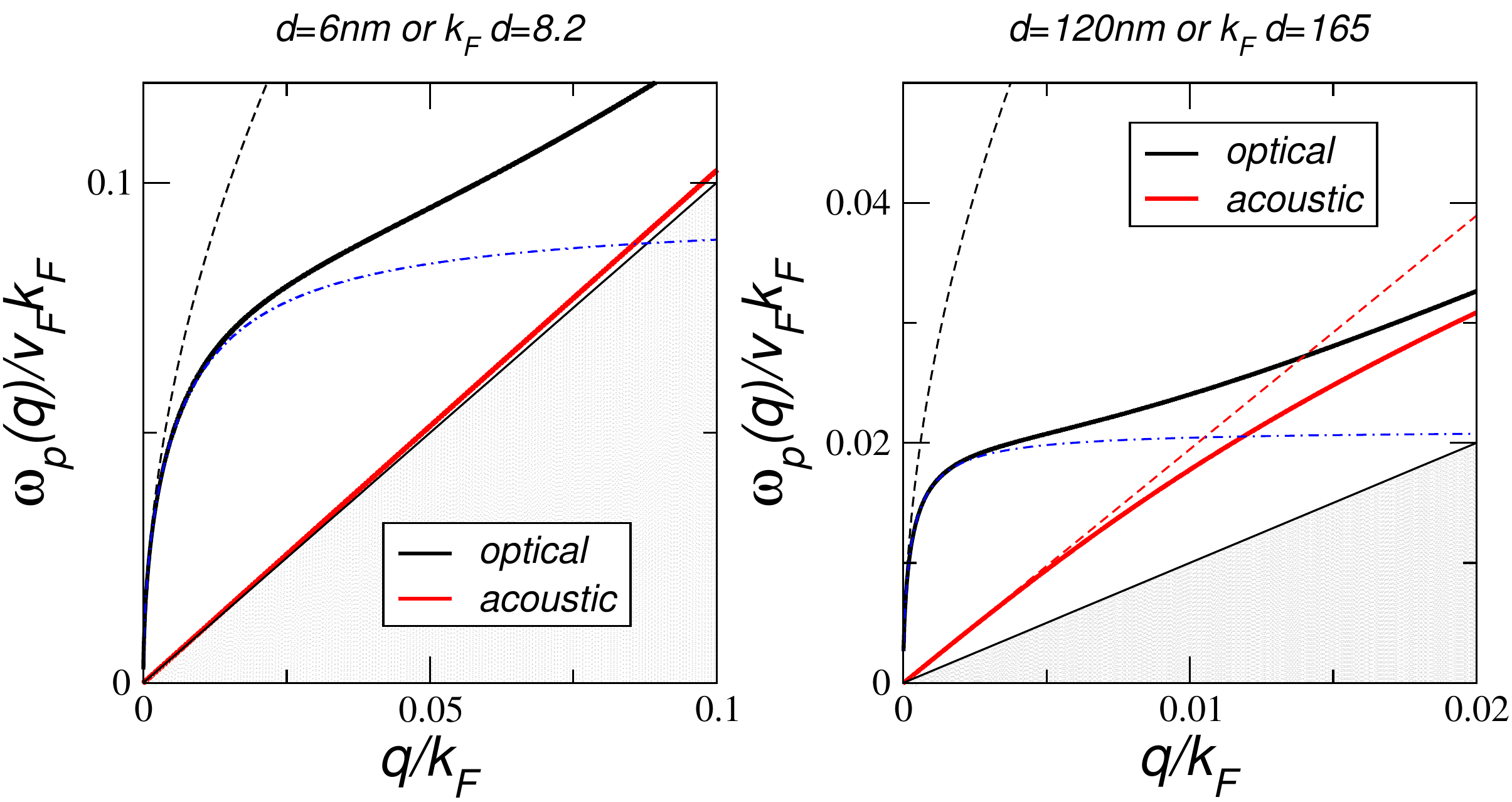}
\caption{(color online): Numerical solutions of the optical (black) and acoustic (red) mode contrasted with the long wavelength approximations of Eq. (\ref{OpticalMode}) (black dashed), (\ref{AcousticModeTwo}) (red dashed) and Eq. (\ref{OpticalModeTwo}) (blue dotted-dashed) for TI-slab width $d=6$nm (left) and $d=120$nm (right). We also show the region of intraband excitations bounded by $\omega=v_Fq$ (grey).}
  \label{figure1}
\end{figure}

In Fig. \ref{figure1}, we show the optical (black) and acoustic (red) plasmonic mode for two slab widths $d=6$nm ($k_Fd=8.2$) (left) and $d=120$nm ($k_Fd=165$). The numerical dispersions are compared to the analytical formulas for the in-phase mode of Eq. (\ref{OpticalMode}) (black dashed line) and the out-of-phase mode of Eq. (\ref{AcousticModeTwo}) (red dashed line). Also shown is the analytical formula of Eq. (\ref{OpticalModeTwo}) as blue dotted-dashed line. We also indicated the regime of intraband excitations bounded by $\omega=v_Fq$ as grey area.

For $d=6$nm, the out-of-phase mode is close to the intraband continuum and Eq. (\ref{AcousticModeTwo}) almost lies on top of the exact solution. But for a larger slab width $d=120$nm, the acoustic mode is already well-separated from $\omega=v_Fq$. Regarding the optical mode, Eq. (\ref{OpticalMode}) can only be applied for small momenta, but Eq. (\ref{OpticalModeTwo}) matches well for $dq\lsim0.2$ which is the upper bound in typical experiments.

\begin{figure}
\centering
  \includegraphics[width=0.99\columnwidth]{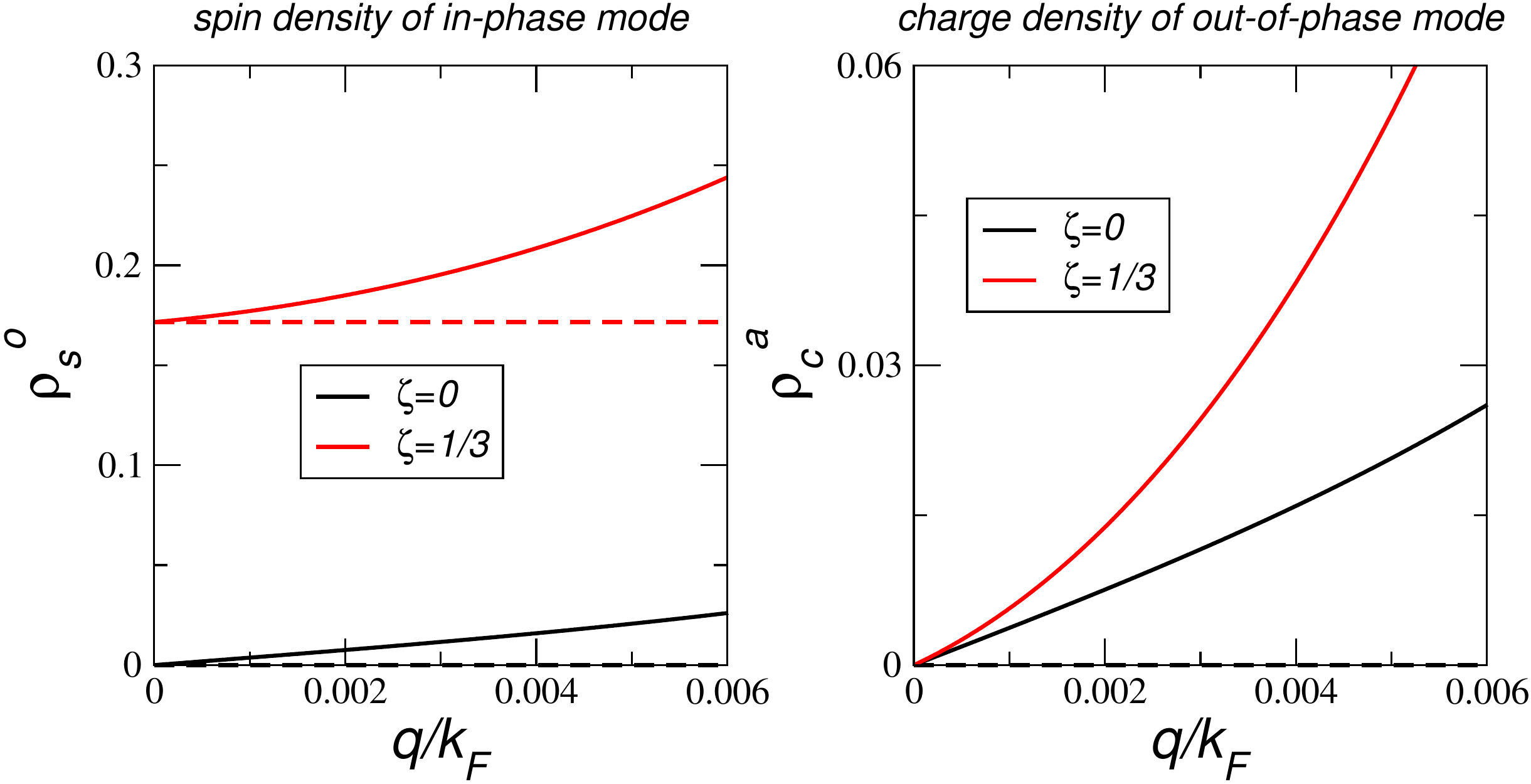}
\caption{Spin density of the optical mode $\rho_s^o$ (left) and charge density of the acoustic mode $\rho_c^a$ (right) for $\zeta=0$ (black) and $\zeta=1/3$ (red) - both normalized with $d=120$nm and $n=3\times10^{13}$cm${}^{-2}$. The analytic results of Eqs. (\ref{OpticalCharge}) and (\ref{AcousticSpin}) are shown as dashed lines (Figure adapted from Ref. \cite{Stauber13}).}
  \label{figureSpinDensity}
\end{figure}

In Fig. \ref{figureSpinDensity}, we discuss the validity of the spin-charge separation beyond the long-wavelength approximation of Eqs. (\ref{OpticalCharge}) and (\ref{AcousticSpin}) (full vs. dashed). The total spin density of the optical mode, $\rho_s^o=|S_\perp^T+S_\perp^B|/(|S_\perp^T|+|S_\perp^B|)$, and the total charge density of the acoustic mode, $\rho_c^a=|\rho^T+\rho^B|/(|\rho^T|+|\rho^B|)$ is shown. For $\zeta=0$ (equal densities on the two surfaces), we find $\rho_s^o=\rho_c^a$ and spin-charge separation is maintained up to $qk_F\lsim5\times10^{-3}$. Wavenumbers in typical experiments are limited by $qk_F\lsim0.001$. We thus expect that spin-charge separation should be observable in this range. For different electronic densities $\zeta\equiv(n_T-n_B)/(n_T+n_B)=1/3$, the in-phase mode is not only charge-like, but also spin-like, although suppressed when compared to the single-layer. The acoustic mode, nevertheless, remains spin-like for the experimentally relevant regime with $\rho_c^a\lsim1$\%.

For equal top and bottom densities $\zeta=0$ and same top and bottom dielectric constants $\epsilon_T=\epsilon_B$, there is perfect spin-charge separation independent of the wave number. Suspended or encapsulated TI-samples should thus display perfect decoupling of the two modes.

\subsection*{Spin-charge separation in nanoribbons}
We now consider the electronic motion confined by a quasi-one dimensional nanowire and assume that the local approximation of the response function is well justified due to the large enough ribbon width $d$. Thus, only the Coulomb interaction must be modified and the longitudinal propagator will give rise to plasmons which disperse as $q'\sqrt{-\ln(q'd)}$.\cite{Brey07,Stauber14} 
The optical mode is then obtained as
\begin{equation}
\omega^2=(\alpha_d v_F^2k_Fd/\pi)q^2\ln(\sqrt{e}/(qd))\;. 
\end{equation}
Considering the logarithmic correction only as an additional factor, not only the acoustic, but also the optical mode shows a linear dispersion.  

Spin-charge separation and collective excitations with linear dispersion are the characteristics of the Tomonaga-Luttinger phenomenology for one-dimensional electron systems. Choosing the width and the length of the ribbon as $a\approx100$nm and $L\approx 10\mu$m, we obtain $v_c\approx10v_F$ in the long-wavelength limit. For the sound (spin) velocity we set $v_s\approx v_F$, valid for small TI slab widths.This would correspond to one-dimensional interacting electrons with Luttinger liquid parameter $K\approx0.1$. By varying the ribbon width to $a\approx10$nm, one can reach $K\approx0.2$, thus being able to tune the effective interaction. 

A helical Luttinger liquid in topological insulator nanowires was recently discussed in Ref. \cite{Egger10} and offers a microscopic theory for the observed spectrum. Including Rashba spin-orbit coupling in the microscopic model leads to a hybridisation of spin- and charge-like excitations.\cite{Moroz00} This is similar to what is predicted in the case of unbalanced charge densities of the thin slab that can be induced by an electric field similarly to the emergence of the Rashba spin-orbit coupling. 

\subsection*{Possible experimental observation}
In Ref. \cite{Wu15}, it was  demonstrated that Cu$_{0.02}$Bi$_2$Se$_3$
can be described by a single TSS component without the mixture of a 2DEG depletion layer or free bulk charges. Purely spin- and charge-like collective oscillations should thus be observable in a thin slab of this topological insulators.

Let us discuss possible experimental consequences. In Ref. \cite{Raghu10}, it was proposed to generate a transverse spin wave by a transient spin grating consisting of two orthogonally polarized non collinear incident beams. This would work for a very thick TIs, but making the sampler width thiner, spin-charge separation will lead to a purely spinless collective optical mode and the proposed coupling mechanism would not be active, anymore. 

The consequences for the acoustic mode might be more interesting. On the right hand side of Fig. \ref{figure1}, it is shown that this mode is well separated from the particle-hole continuum for wider slabs and should hence be observable, see also Ref. \cite{Ameen17} for a discussion. Further, it could be effectively excited by neutrons or neutral atoms/molecules that carry a spin, such as $H_2$.  

\section*{Influence of depletion layer}
Let us now investigate the influence of the depletion layer on the plasmonic excitations of a TI. This will be done by comparing the predictions of our theory with the experimental data. We will first comment on the terahertz response of patterned as well as of unpatterned samples. We will then briefly discuss other experiments involving magnetic fields as well as time-dependent spectroscopy.  
\subsection*{Terahertz response of patterned samples}
Due to their large momentum, longitudinal plasmons cannot interact with propagating electromagnetic radiation. In order to overcome the momentum mismatch, the sample is patterned with a periodic sub wavelength structure. This has been done for graphene showing plasmon resonances with remarkably large oscillator strengths at room temperature.\cite{Ju12,Yan13}

A patterned superlattice can also be used to measure plasmonic resonances of topological insulators.\cite{DiPietro13,Autore15a,Autore15b,Post15,Sim15,Autore16} Usually, Eq. (\ref{OpticalMode}) can be used to describe the plasmonic dispersion which is valid in the long-wavelength limit independent of the dielectric constant of the TI. But the dielectric constant of the TI is comparably large such that the first order correction of Eq. (\ref{OpticalModeTwo}) can yield a sizable corrections at larger inplane momenta. 

For the comparison, we will use parameters for the Bi${}_2$Se${}_3$ family with $v_F=6\times10^5$m/s leading to $\alpha_d=3.7$.\cite{DiPietro13} The chemical potential is obtained from the total electron density of $n=n_T+n_B=3\times10^{13}$cm${}^{-2}$.\cite{Bansal12,DiPietro13} Furthermore, Bi${}_2$Se${}_3$ was grown on a sapphire substrate (Al${}_2$O${}_3$) and we set the lower and upper dielectric constant $\epsilon_B=10$ and $\epsilon_T=1$ (air), respectively, as in Ref. \cite{DiPietro13}. 

The value of the topological insulator needs more discussion. The static value of topological insulators is approximately given by $\epsilon_{TI}=100$,\cite{Profumo12} which is composed of a contribution coming from $\alpha$-phonons and a constant high-frequency background. For frequencies above the phonon frequency $\omega>\omega_{phon}^{\alpha}$=1.85THz, the phonon contribution is almost exhausted and $\epsilon_{TI}^\infty\approx30$.\cite{Stordeur92,Nechaev13} 

We are now in the position to plot the optical mode of Eq. (\ref{OpticalModeTwo}) for slab widths $d=60$nm (black) and $d=120$nm (red) with $\epsilon_{TI}=100$ (solid lines) and $\epsilon_{TI}=30$ (dashed lines), shown in Fig. \ref{Compare}. For the small wave numbers shown here, the analytic curves are indistinguishable from the exact numerical solution, but already depend on the slab width $d$. Notice that they lie below the experimental data which were taken for two different slab widths $d=60$nm represented as circles and $d=120$nm represented as squares. 

Eq. (\ref{OpticalMode}) is shown as dashed line and the good agreement with the experimental data is still missing an explanation. In the following, we will discuss two possible extensions to the above model in order to explain the experiments. The first extension will include the effect of a 2D depletion layer maintaining the model with constant dielectric media. As second extension, we will use a frequency dependent dielectric function as discussed in Sec. \ref{Macroscopic}.   

\subsection*{Including 2D depletion layer}
From the above discussion we see that the spectral weight coming from the Dirac electrons is not enough to describe the experimental data for large $q$. But there is also a two-dimensional spin-degenerate electron gas (2DEG) that can add additional spectral weight.\cite{Bianchi10} Indeed, two channels, both conducting, were found, one corresponding to the topologically protected Dirac electrons and the other to the 2DEG that appears due to band bending effects.\cite{Bansal12}  

Including the depletion layer would also increase the dimension of the considered response matrix by a factor of two. However, these additional modes do not change our theory as outlined in Ref. \cite{Stauber13}. In what follows, we will thus not consider particle exchange between the Dirac carriers and depletion layer and also the emerging charge-less acoustic modes. The total charge response is then obtained as the sum of the response of the Dirac and Schr\"odinger electrons, 
\begin{align}
\label{local2}
\chi_{\rho\rho}^{tot}=\chi_{\rho\rho}^{T/B}+\chi_{\rho\rho}^{2DEG}\;.
\end{align}
The charge response of a 2DEG was first derived in Ref. \cite{Stern67} but for our purposes, the local approximation is sufficient. We thus have
\begin{equation}
\label{2DEG}
\chi_{\rho\rho}^{2DEG}=\frac{\mu^{2DEG}}{\pi\hbar^2}\frac{q^2}{\omega^2}\;.
\end{equation}

We have separate Fermi energies for Dirac and Schr\"odinger fermions and
$\mu^{Dirac}$ and $\mu^{2DEG}$ are measured with respect to the Dirac point and wirht respect to the bottom of the 2DEG band, respectively. We can thus use the initial formula of Eq. (\ref{OpticalModeTwo}), but with a different chemical potential: 
\begin{align}
\mu\to\mu^{Dirac}+4\mu^{2DEG}\;.
\end{align}
The extra factor 4 in front of $\mu^{2DEG}$ comes from comparing Eqs. (\ref{local}) and (\ref{2DEG}).

\begin{figure}
\centering
  \includegraphics[width=0.99\columnwidth]{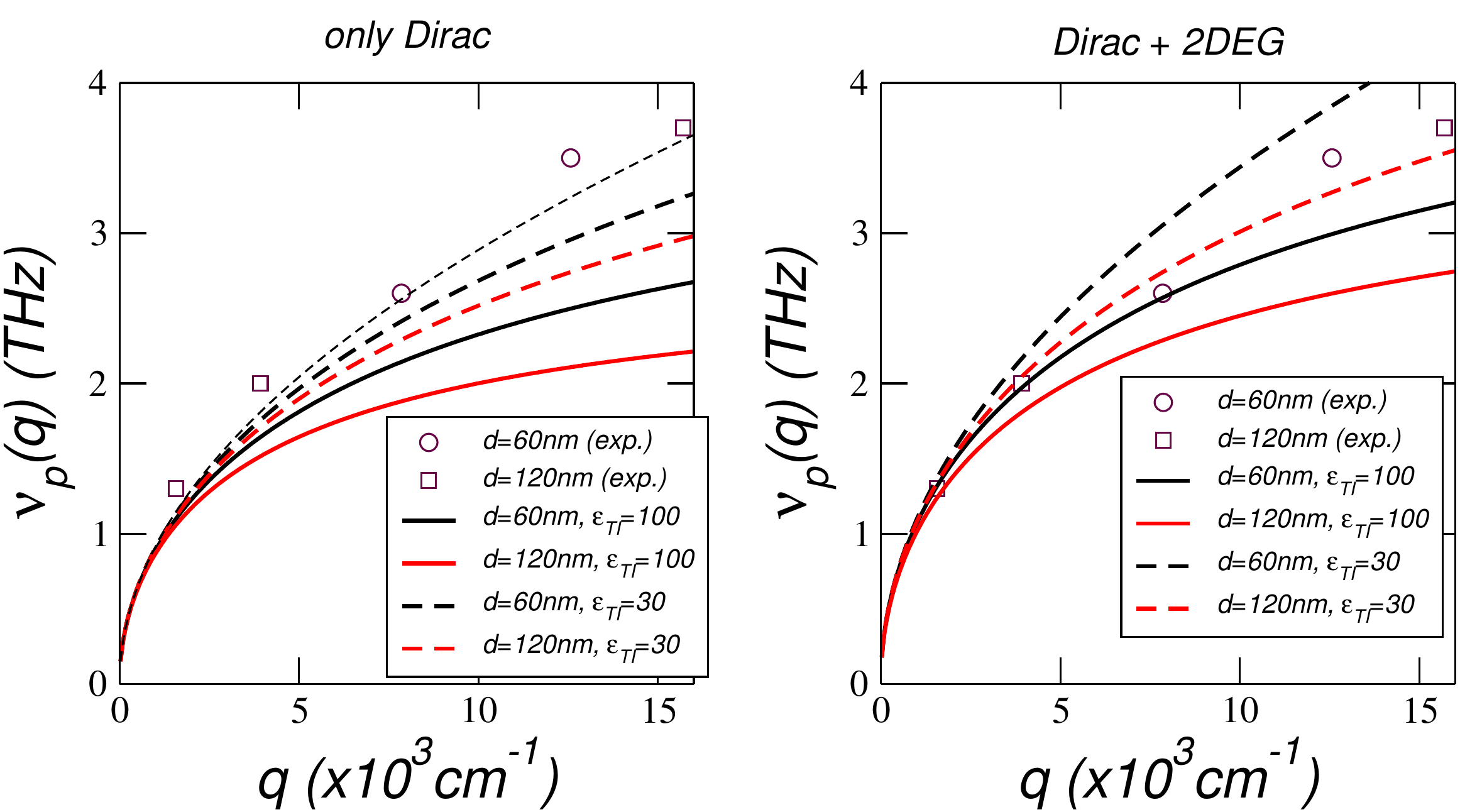}
\caption{Comparison of the experimental data of Ref. \cite{DiPietro13} (symbols) to the analytical result of Eq. (\ref{OpticalModeTwo}). Left: Including only Dirac Fermions using a different slab width $d=60$nm (black), $d=120$nm (red) for different dielectric constants $\epsilon_{TI}=100$ (full lines) and $\epsilon_{TI}=30$ (dashed lines). Also shown the result of Eq. (\ref{OpticalMode}) (dashed). Right: Adding to the Dirac Fermions the contribution of the 2DEG with  $d=60$nm (black), $d=120$nm (red) for $\epsilon_{TI}=100$ (full lines) and $\epsilon_{TI}=30$ (dashed lines).}
  \label{Compare}
\end{figure}

Using the sheet density, we have $\mu^{Dirac}=542$meV, but using the slightly curved energy dispersion would give the slightly smaller value $\mu=450$meV.\cite{Bansal12} In any case, with $\mu^{2DEG}=60$meV,\cite{Bansal12} we obtain good agreement with the experimental for small wave numbers $q\lsim10^4$cm${}^{-1}$, shown in Fig. \ref{Compare} (right hand side). However, uncertainties upon the initial parameters remain which might result in quantitative (but not qualititive) differences. In the same figure, we plot the plasmon frequencies $\nu_p=\omega_+/2\pi$ for two different widths $d=60$nm (black line) and $d=120$nm (red line). We also show the results for two different dielectric TI-constants $\epsilon_{TI}=100$ (full lines) and $\epsilon_{TI}=30$ (dashed lines). 

For larger momenta $q>10^4$cm${}^{-1}$, the results are blue shifted, but the dipole-dipole interaction between the patterned nano wires should add an additional red-shift, at least for samples with small periodicities;\cite{Nikitin12,Christensen12} this shift can be as large as 20\% for small periodicities of $2\mu$m with equal spacing.\cite{Strait13} Considering the frequency dependence of the substrate dielectric function $\epsilon_B$ could lead to effectively smaller values $\epsilon_B(\omega)<10$.\cite{Kim09}  

Let us finally note that the inclusion of a Drude term that corresponds to free bulk electrons (and not to the topological protected surface states) will lead to a considerable increase of the spectral weight. This term was analysed in Ref. \cite{Deshko16} using Eq. (\ref{OpticalModeTwo}).

\subsection*{Terahertz response in unpatterned samples}
In Ref. \cite{Valdes12}, the Drude weight of the topologically protected surface states (TSS) of Bi$_2$Se$_3$ was obtained from fitting the terahertz response to the macroscopic model assuming the dielectric constant as a sum of two Drude-Lorentz oscillators, Eq. (\ref{Drude-Lorentz}). The conductivity data could be fitted {\em without} the assumption of an additional depletion layer. Nevertheless, in Ref. \cite{Wu15} the same group confirmed the existence of an additional contribution to the Drude weight by magneto-THz measurements.\cite{Wu15} The obtain the spectral weight of the TSS  which amounts to 90\% of the total spectral weight. From the transport analysis of \cite{Bansal12}, we obtain a percentage of 70\%, as outlined in the previous section. By slightly changing the parameters, e.g., measuring the Fermi energy differently, we would obtain a percentage of around 80\%, but still, a discrepancy between the two experiments/interpretations remains and we will try to resolve it in the following. 

The analysis of Ref. \cite{Wu15} is based on the f-sum rule and one of the virtues of the f-sum rule is that it is also valid for interacting systems since it relies on charge conservation. Integrating the optical conductivity over the whole spectral range is then related to the Drude weight which is independent of the interaction in a Galilean invariant system, see Eq. (\ref{SumRuleD}) of the appendix. Nevertheless, this is not the case for Dirac systems anymore, and electron-electron interactions modify the Drude weight in a non-trivial way. Moreover, the Drude weight of the interacting system is larger than the Drude weight of the non-interacting system.\cite{Abedinpour11,Levitov13,Stauber17} 

The experimental result of 90\% spectral weight of the TSS thus includes possible interaction effects. Neglecting those, as done by our theory above, would lower the result to become closer to the 70\%-80\%, first obtained in Ref. \cite{Stauber13}. Obviously, the screening effect of the TI-substrate is large and will suppress interaction effects. On the other hand, there is only one electronic flavor which will decrease the intrinsic screening compared to graphene by a factor of four. To summarize, the spectral weight of Bi$_2$Se$_3$ should be predominately composed of Dirac Fermions emerging from the TSS and ranging between 80\%-90\%.

\subsection*{Further experiments}
In Ref. \cite{Autore16}, the quantum phase transition in the (Bi$_{1-x}$In$_x$)$_2$Se$_3$ topological insulator can be detected by the change of the plasmon scattering rate as function of $x$. Only in the topologically protected phase with $x>x_{cr}$, there are TSS and the scattering rate is mainly given by the Dirac plasmon. In the normal phase, only the usual 2DEG plasmon composed by massive (Schr\"odinger) electrons are present which give rise to a considerably larger scattering rate. 

In Ref. \cite{Sim15} the modulation depth of the 3D topological insulators Bi$_2$Se$_3$ was investigated via an optical pump-THz probe spectroscopy. It was shown that the  interplay between topologically protected surface states, 2DEG electrons and semiconductor bulk electrons as well as the plasmon-phonon coupling are responsible for an unprecedented modulation depth of 2400\%. 

Both examples indicate the existence and importance of the 2DEG depletion layer. It could further give rise to the existence of interband plasmons for which a departure from pure Dirac carriers is necessary. For genuine interband plasmons, the Drude weight $D$ is zero and the collective oscillations are formed by the light-induced electronic and hole densities in the conduction and valence band, respectively. For pure Dirac electrons, the response function is negative for all wavenumbers $q$ and energies $\omega$ such that the condition $\epsilon=0$ can never be satisfied. Only the admixture of Schr\"odinger electrons might cause the the dielectric function to become zero which is the case for HgTe\cite{Juergens14b} or for twisted bilayer graphene.\cite{StauberNJP13,Stauber16}

\section*{Plasmon-phonon coupling}
Let us now discuss the influence of a frequency dependent dielectric function for the topological insulator as well as for the bulk substrate. By this, we can include the coupling of the plasmon modes to the surface phonons of the TI. The model for the dielectric function of the topological insulator is given by the Drude-Lorentz model, Eq. (\ref{Drude-Lorentz}). The parameters are taken from Ref. \cite{DiPietro13} for the patterned sample with $W=8\mu$m at temperature $T=6K$. The model for sapphire-substrate is taken from Ref. \cite{Roberts62} even though assuming a constant dielectric medium $\epsilon_B$ is well justified in the considered frequency range.

Due to the damping terms of the phonon modes, the dielectric constant becomes complex and the plasmonic excitations turn into resonances. The calculation of the energy loss function would thus be appropriate.\cite{Stauber12} Still, we can maintain the above discussion of the plasmonic dispersion by solving only the real part of Eq. (\ref{det}). This will yield multiple plasmonic branches, but only the ones corresponding to the smallest imaginary part would carry most of the spectral weight. These shall be discussed in the following.  

\begin{figure}
\centering
  \includegraphics[width=0.99\columnwidth]{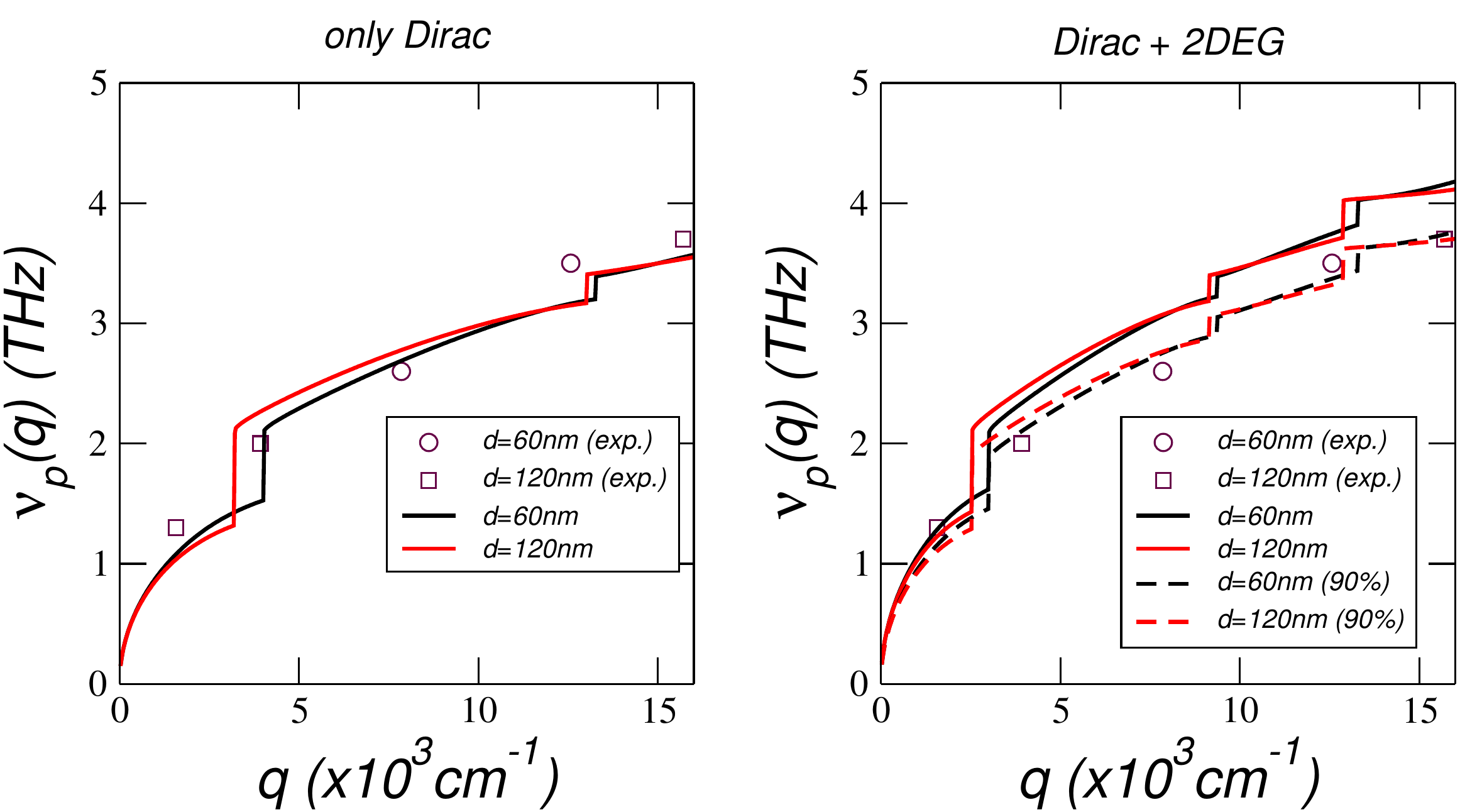}
\caption{Comparison of the experimental data of Ref. \cite{DiPietro13} (symbols) with the optical mode of the real part of Eq. (\ref{det}). Right: Including only Dirac Fermions with $d=60$nm (black), $d=120$nm (red) for $\epsilon(\omega)$ at $T=6$K as given in Fig. \ref{Dielectric}. Right: Same as on the left hand side including both Dirac Fermions and 2DEG. Dashed curves correspond to a phenomenological red-shift due to dipole-dipole interaction between the patterned nano wires.}
  \label{CompareBulk}
\end{figure}

In Fig. \ref{CompareBulk}, we show the resulting plasmonic dispersion of the optical mode for two different slab widths $d=60$nm and $d=120$nm (red) with (right panel) and without (left panel) the contribution of the 2DEG. There are discontinuities at the two phonon frequencies which are related to the hybridisation between the plasmon and the phonon modes. These hybridisations were also observed in a similar context in Ref. \cite{Scholz12} or for graphene on a SiO$_2$-substrate.\cite{Yan13} Moreover, the plasmon frequency of the wider slab is usually above the one with of the narrower slab width. This is because for frequencies above the phonon frequency $\omega>\omega_{phon}^{\alpha}$=1.85THz, the real part of the dielectric function becomes negative. 

As noted above, the dipole-dipole interaction between the patterned nano wires should lead to an additional red-shift compared to the analytic curves for samples with small periodicities.\cite{Nikitin12,Christensen12} This shift grows for increasing wave number $q$ and can become as large as 10\% for small periodicities of $2\mu$m with equal spacing.\cite{Strait13} The resulting curves are shown as dashed lines on the right hand side of Fig. \ref{CompareBulk}. The comparison to the experimental data would thus suggest the existence of the 2DEG to account for the additional spectral weight.  
 
\begin{figure}
\centering
  \includegraphics[width=0.49\columnwidth]{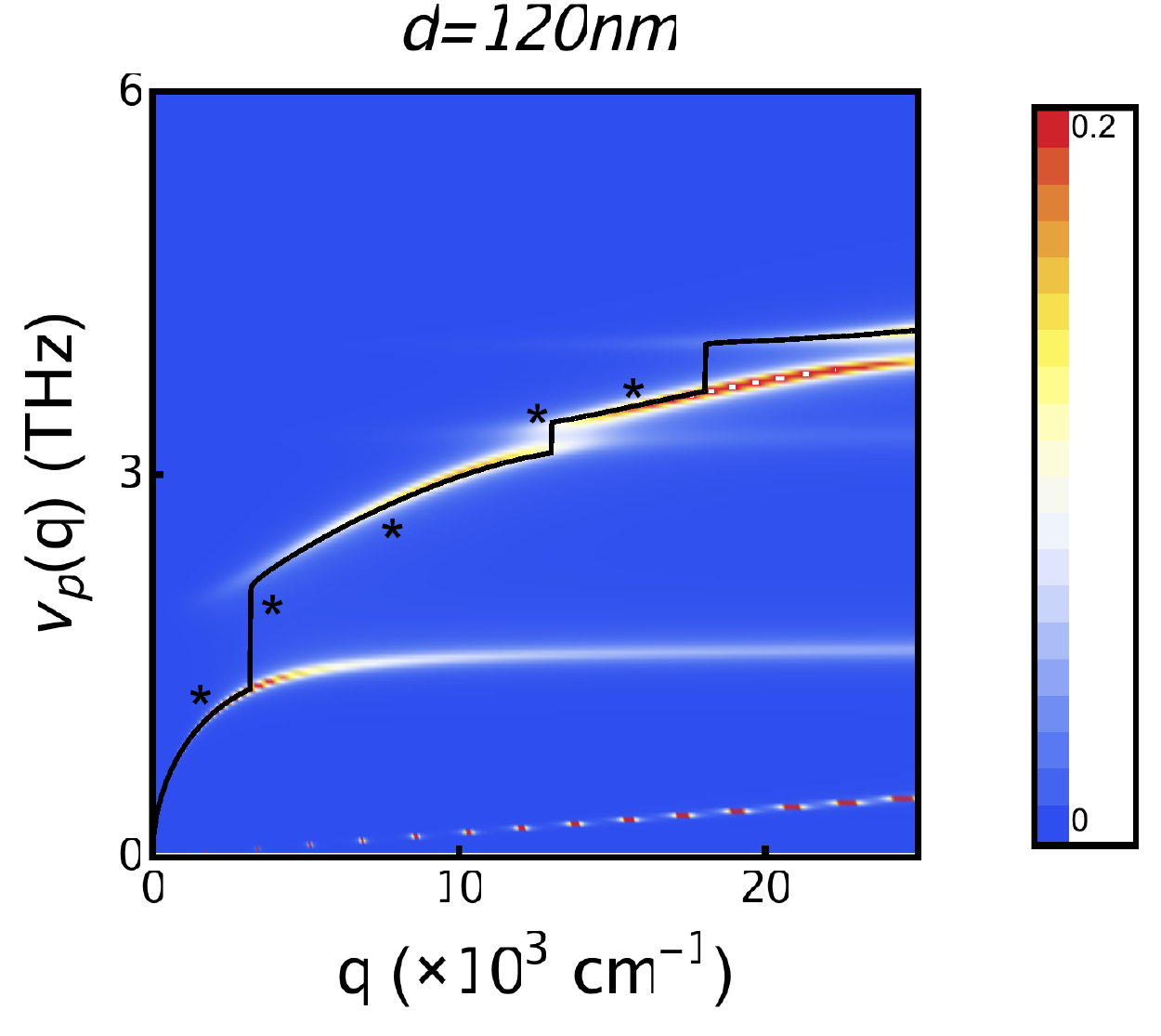}
  \includegraphics[width=0.49\columnwidth]{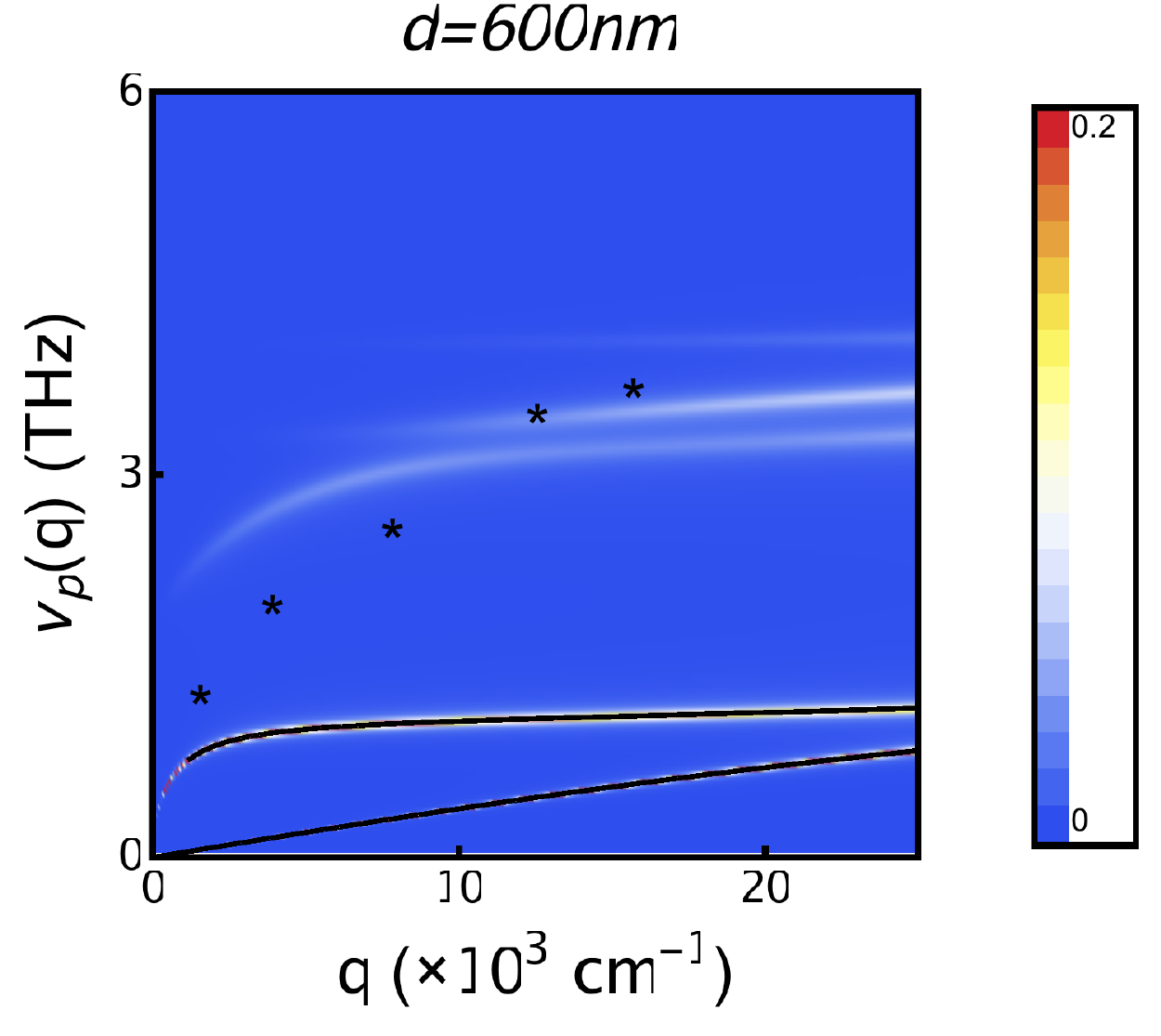}
\caption{The energy loss function $S(q,\omega)$ in units of $\mu/\hbar^2$ for two different slab widths $d=120$nm (left) and $d=600$nm (right). Also shown is the experimental data of Ref. \cite{DiPietro13} (stars) and the solution for the zeros of the real part of the dielectric function with large spectral weight as discussed in Fig. \ref{CompareBulk} (solid lines).}
  \label{DensPlot}
\end{figure}

Let us finally discuss the plasmonic spectrum in terms of the electron energy loss function defined as
\begin{align}
S(q,\omega)=-\frac{1}{e^2}\;\rm{Im}\;\rm{Tr}\;\chi_{jj}(q,\omega)\;,
\end{align}
where $\chi_{jj}=\frac{\omega^2}{q^2}\chi_{\rho\rho}$ is the effective $2\times2$ current-current response function.\cite{Stauber12} On the left hand side of Fig. \ref{DensPlot}, we display the density plot of the loss function for the slab width $d=120$nm. The hybridization between the (surface) $\alpha$-phonon mode and the plasmon mode is clearly seen and should also be experimentally observable (see below). Also the acoustic mode carries some spectral weight that is detectable. 

On the right hand side of Fig. \ref{DensPlot}, the loss function is shown for a wider slab with $d=600$nm. Now, most of the spectral weight has been transferred to the quasi-localized phonon mode which is nevertheless strongly renormalized by the hybridization with the plasmon mode. Together with the acoustic mode, these are the two modes which carry considerable spectral weight. They are also considerably red-shifted compared to the case of a static dielectric medium, i.e., if we would have, e.g., simply used $\epsilon_{TI}=30$. 

The hybridization of phonon and plasmon modes has recently been observed displaying the same features of red-shifted energy and long life-time.\cite{Jia17} We expect the long life-time to be induced by the long-lived phonon mode and {\it not} by the topologically protected surface modes. In any case, the strong dependence of the plasmonic resonances on the slab thickness would also open up new possibilities of engineering the these TI plasmon modes.   

\section*{Conclusions}
We have discussed plasmonic excitations in thin slabs of topological insulators. The electronic properties of the 2D-surface of a topological insulator is characterised by topologically protected surface states (TSS) and well approximated by Dirac fermions. The Drude weight is then given by $D=\frac{e^2v_F}{\hbar}\sqrt{\frac{n}{4\pi}}$ which defines the plasmon dispersions for low $q$. For larger $q$, there are deviations from this behaviour but which are not yet at experimental reach using micro-slabs.   

We have also pointed out that the plasmonic excitations in a thin slabs of TIs display spin-charge separation, i.e., the in-phase modes are purely charge-like. This might be the reason why the associated spin-character of the collective modes - predicted for a single layer - have not been detected in thin slabs, so far (large slab width would induce additional noise from the bulk). The spin-charge separation of TI also suggests a relation to thin nano-wires that behave as helical Luttinger liquids.\cite{Egger10} 

As another result, we found a novel thickness dependence of the plasmonic modes valid even in the regime of small wave numbers. In the special case of equal electronic densities at the two layers, the formula is also valid for general quasi-2D systems deduced by using the Keldych potential.\cite{Bondarev17} From that, we argued that in addition to the response of the Dirac carriers, also the two-dimensional electron gas (2DEG) of the depletion layer underneath the TI surface contributes to the spectral weight of the plasmonic resonances. An interplay of surface and Dirac plasmons in topological insulators was observed by several experimental groups in the case of Bi${}_2$Se${}_3$.\cite{Sim15,Politano15,Wu15,Autore16} 

Finally, we discussed the plasmonic excitations including the $\alpha$ and $\beta$ phonons of the bulk TI. We observe a hybridisation of the surface phonon and plasmon modes and also a transfer of spectral weight to lower lying modes when the slab width is increased up to $d=600$nm. There is also an increase of the life time of this hybrid mode induced by the long-lived phonon mode as observed in Ref. \cite{Jia17} and a strong tuneability.
\section*{Acknowledgments} 
This work has been supported by Spain's MINECO under grants FIS2012-37549-C05-03, FIS2014-57432-P, FIS2015-64654-P, and FIS and by the Comunidad Madrid under S2013/MIT-3007 MAD2D-CM.

\section*{Appendix: Derivation of the f-sum rule}
The f-sum rule can be deduced from the Kramers-Kronig relation for the paramagnetic current-current correlation function:
\begin{align}
\frac{1}{\pi}\mathcal{P}\int_{-\infty}^{\infty}d\omega' \frac{\im\chi^P(\omega')}{\omega'-\omega}=\re\chi^P(\omega)
\end{align}
where $\mathcal{P}$ denotes the principle value. Above, we defined $\chi^P=\chi^{\alpha,\alpha}$ with 
\begin{equation}
\label{LinearResponse}
\chi^{\alpha,\beta}(\omega)=-\frac{\I}{\hbar}\int_0^\infty e^{\I\omega t}\langle [j^\alpha(t),j^\beta(0)]\rangle\;,
\end{equation}
and the paramagnetic current $j^\alpha$ in direction $\alpha$.

For $\omega=0$ and the relation of $\chi^P$ to the electric conductivity, $\chi^P(\omega)=-\I\omega\sigma^P(\omega)$, this reads
\begin{align}
\mathcal{P}\int_{-\infty}^{\infty}d\omega'\re\sigma^P(\omega')=\pi\re\chi^P(\omega=0)\;.
\end{align}
The full real part of the conductivity is usually decomposed into a regular part $\re\sigma_{reg}^P$ as well as a contribution containing the delta function
\begin{align}
\re\sigma^P(\omega)=\pi\re\chi^P(\omega)\delta(\omega)+\re\sigma_{reg}^P(\omega)\;.
\end{align}
The sum-rule then reads
\begin{align}
\int_{-\infty}^{\infty}d\omega\re\sigma^P(\omega)=0\;.
\end{align} 

The physical current also contains a diamagnetic contribution and for a tight-binding model this contribution is proportional to the kinetic energy.\cite{Scalapino93} This implies a constant diamagnetic response independent of $\omega$ with $\j^{dia}(\omega)=-\chi^{dia}\A(\omega)$ and $\chi^{dia}>0$. The full conductivity thus reads 
 \begin{align}
\re\sigma(\omega)=\re\sigma^P(\omega)+\pi\chi^{dia}\delta(\omega)\;
\end{align}
and the corresponding sum rule
\begin{align}
\int_{-\infty}^{\infty}d\omega\re\sigma(\omega)=\pi\chi^{dia}\;.
\end{align} 
With the particle density $n$ and electron mass $m$, $\chi^{dia}=\frac{e^2n}{m}$ and the typical version of the f-sum rule is thus derived.

The charge stiffness or Drude weight is defined as $D=\pi[\chi^{dia}+\re\chi^P(\omega=0)]$. For a Galilean invariant (i.e., translationally invariant) system, we have $j^\alpha(t)=$const. and thus $\chi^P(\omega)=0$. This then yields the typical Drude weight $D=\frac{\pi e^2n}{m}$ and the following version of the sum rule:
\begin{align}
\label{SumRuleD}
\int_{0}^{\infty}d\omega\re\sigma(\omega)=\frac{D}{2}\;.
\end{align}  

Note that the above expression is valid for any dimension of the underlying system. For three-dimensional systems often the relation to the plasmon frequency $\omega_p^2=\frac{e^2n}{\epsilon_0m}$ is discussed, i.e.,
\begin{align}
\label{SumRuleD_Dirac}
\frac{2}{\pi\epsilon_0}\int_{0}^{\infty}d\omega\re\sigma(\omega)=\omega_p^2\;.
\end{align} 
Relating now $\omega_p^2d$ with the Drude weight in two dimensions, i.e., $\omega_p^2d=\frac{e^2n}{\epsilon_0m}$, i.e.,
 is the procedure how to obtain the density of the localized TSS via the sum rule.\cite{Valdes12,Wu15} The Fermi wave vector $k_F$, the effective mass (in case of the Dirac Fermion this is defined as $m^*=\hbar k_F/v_F$) as well as the resulting Fermi energy $E_F$ from the assumed dispersion can thus be read of the extinction spectrum. For a general discussion on the sum rule based on a multi-band tight-binding model including the wave number $q$, see Ref. \cite{Stauber10b}.

\bibliography{topological}

\providecommand{\latin}[1]{#1}
\providecommand*\mcitethebibliography{\thebibliography}
\csname @ifundefined\endcsname{endmcitethebibliography}
  {\let\endmcitethebibliography\endthebibliography}{}
\begin{mcitethebibliography}{69}
\providecommand*\natexlab[1]{#1}
\providecommand*\mciteSetBstSublistMode[1]{}
\providecommand*\mciteSetBstMaxWidthForm[2]{}
\providecommand*\mciteBstWouldAddEndPuncttrue
  {\def\EndOfBibitem{\unskip.}}
\providecommand*\mciteBstWouldAddEndPunctfalse
  {\let\EndOfBibitem\relax}
\providecommand*\mciteSetBstMidEndSepPunct[3]{}
\providecommand*\mciteSetBstSublistLabelBeginEnd[3]{}
\providecommand*\EndOfBibitem{}
\mciteSetBstSublistMode{f}
\mciteSetBstMaxWidthForm{subitem}{(\alph{mcitesubitemcount})}
\mciteSetBstSublistLabelBeginEnd
  {\mcitemaxwidthsubitemform\space}
  {\relax}
  {\relax}

\bibitem[Basov \latin{et~al.}(2016)Basov, Fogler, and Garc{\'\i}a~de
  Abajo]{Basov16}
Basov,~D.~N.; Fogler,~M.~M.; Garc{\'\i}a~de Abajo,~F.~J. Polaritons in van der
  Waals materials. \textbf{2016}, \emph{354}, 195\relax
\mciteBstWouldAddEndPuncttrue
\mciteSetBstMidEndSepPunct{\mcitedefaultmidpunct}
{\mcitedefaultendpunct}{\mcitedefaultseppunct}\relax
\EndOfBibitem
\bibitem[Low \latin{et~al.}(2017)Low, Chaves, Caldwell, Kumar, Fang, Avouris,
  Heinz, Guinea, Martin-Moreno, and Koppens]{Low17}
Low,~T.; Chaves,~A.; Caldwell,~J.~D.; Kumar,~A.; Fang,~N.~X.; Avouris,~P.;
  Heinz,~T.~F.; Guinea,~F.; Martin-Moreno,~L.; Koppens,~F. Polaritons in
  layered two-dimensional materials. \emph{Nat Mater} \textbf{2017}, \emph{16},
  182--194\relax
\mciteBstWouldAddEndPuncttrue
\mciteSetBstMidEndSepPunct{\mcitedefaultmidpunct}
{\mcitedefaultendpunct}{\mcitedefaultseppunct}\relax
\EndOfBibitem
\bibitem[Hsieh \latin{et~al.}(2009)Hsieh, Xia, Qian, Wray, Dil, Meier,
  Osterwalder, Patthey, Checkelsky, Ong, Fedorov, Lin, Bansil, Grauer, Hor,
  Cava, and Hasan]{Hsieh09}
Hsieh,~D. \latin{et~al.}  A tunable topological insulator in the spin helical
  Dirac transport regime. \emph{Nature} \textbf{2009}, \emph{460},
  1101--1105\relax
\mciteBstWouldAddEndPuncttrue
\mciteSetBstMidEndSepPunct{\mcitedefaultmidpunct}
{\mcitedefaultendpunct}{\mcitedefaultseppunct}\relax
\EndOfBibitem
\bibitem[Chen \latin{et~al.}(2009)Chen, Analytis, Chu, Liu, Mo, Qi, Zhang, Lu,
  Dai, Fang, Zhang, Fisher, Hussain, and Shen]{Chen09}
Chen,~Y.; Analytis,~J.; Chu,~J.; Liu,~Z.; Mo,~S.; Qi,~X.; Zhang,~H.; Lu,~D.;
  Dai,~X.; Fang,~Z.; Zhang,~S.; Fisher,~I.; Hussain,~Z.; Shen,~Z. Experimental
  realization of a three-dimensional topological insulator, Bi2Te3.
  \emph{Science} \textbf{2009}, \emph{325}, 178--181\relax
\mciteBstWouldAddEndPuncttrue
\mciteSetBstMidEndSepPunct{\mcitedefaultmidpunct}
{\mcitedefaultendpunct}{\mcitedefaultseppunct}\relax
\EndOfBibitem
\bibitem[Xia \latin{et~al.}(2009)Xia, Qian, Hsieh, Wray, Pal, Lin, Bansil,
  Grauer, Hor, Cava, and Hasan]{Xia09}
Xia,~Y.; Qian,~D.; Hsieh,~D.; Wray,~L.; Pal,~A.; Lin,~H.; Bansil,~A.;
  Grauer,~D.; Hor,~Y.; Cava,~R.; Hasan,~M. Observation of a large-gap
  topological-insulator class with a single Dirac cone on the surface.
  \emph{Nature Phys} \textbf{2009}, \emph{5}, 398--402\relax
\mciteBstWouldAddEndPuncttrue
\mciteSetBstMidEndSepPunct{\mcitedefaultmidpunct}
{\mcitedefaultendpunct}{\mcitedefaultseppunct}\relax
\EndOfBibitem
\bibitem[Stauber \latin{et~al.}(2013)Stauber, G\'omez-Santos, and
  Brey]{Stauber13}
Stauber,~T.; G\'omez-Santos,~G.; Brey,~L. Spin-charge separation of plasmonic
  excitations in thin topological insulators. \emph{Phys. Rev. B}
  \textbf{2013}, \emph{88}, 205427\relax
\mciteBstWouldAddEndPuncttrue
\mciteSetBstMidEndSepPunct{\mcitedefaultmidpunct}
{\mcitedefaultendpunct}{\mcitedefaultseppunct}\relax
\EndOfBibitem
\bibitem[Hasan and Kane(2010)Hasan, and Kane]{Hasan10}
Hasan,~M.; Kane,~C. Colloquium: Topological insulators. \emph{Rev Mod Phys}
  \textbf{2010}, \emph{82}, 3045--3067\relax
\mciteBstWouldAddEndPuncttrue
\mciteSetBstMidEndSepPunct{\mcitedefaultmidpunct}
{\mcitedefaultendpunct}{\mcitedefaultseppunct}\relax
\EndOfBibitem
\bibitem[Qi and Zhang(2011)Qi, and Zhang]{Qi11}
Qi,~X.-L.; Zhang,~S.-C. Topological insulators and superconductors. \emph{Rev.
  Mod. Phys.} \textbf{2011}, \emph{83}, 1057--1110\relax
\mciteBstWouldAddEndPuncttrue
\mciteSetBstMidEndSepPunct{\mcitedefaultmidpunct}
{\mcitedefaultendpunct}{\mcitedefaultseppunct}\relax
\EndOfBibitem
\bibitem[Raghu \latin{et~al.}(2010)Raghu, Chung, Qi, and Zhang]{Raghu10}
Raghu,~S.; Chung,~S.~B.; Qi,~X.-L.; Zhang,~S.-C. Collective Modes of a Helical
  Liquid. \emph{Phys. Rev. Lett.} \textbf{2010}, \emph{104}, 116401\relax
\mciteBstWouldAddEndPuncttrue
\mciteSetBstMidEndSepPunct{\mcitedefaultmidpunct}
{\mcitedefaultendpunct}{\mcitedefaultseppunct}\relax
\EndOfBibitem
\bibitem[Efimkin \latin{et~al.}(2012)Efimkin, Lozovik, and Sokolik]{Efimkin12}
Efimkin,~D.; Lozovik,~Y.; Sokolik,~A. Spin-plasmons in topological insulator.
  \emph{Journal of Magnetism and Magnetic Materials} \textbf{2012}, \emph{324},
  3610 -- 3612\relax
\mciteBstWouldAddEndPuncttrue
\mciteSetBstMidEndSepPunct{\mcitedefaultmidpunct}
{\mcitedefaultendpunct}{\mcitedefaultseppunct}\relax
\EndOfBibitem
\bibitem[Efimkin \latin{et~al.}(2012)Efimkin, Lozovik, and Sokolik]{Efimkin12b}
Efimkin,~D.; Lozovik,~Y.; Sokolik,~A. Collective excitations on a surface of
  topological insulator. \emph{Nanoscale Research Letters} \textbf{2012},
  \emph{7}, 163\relax
\mciteBstWouldAddEndPuncttrue
\mciteSetBstMidEndSepPunct{\mcitedefaultmidpunct}
{\mcitedefaultendpunct}{\mcitedefaultseppunct}\relax
\EndOfBibitem
\bibitem[Ritchie(1957)]{Ritchie57}
Ritchie,~R.~H. Plasma Losses by Fast Electrons in Thin Films. \emph{Phys. Rev.}
  \textbf{1957}, \emph{106}, 874--881\relax
\mciteBstWouldAddEndPuncttrue
\mciteSetBstMidEndSepPunct{\mcitedefaultmidpunct}
{\mcitedefaultendpunct}{\mcitedefaultseppunct}\relax
\EndOfBibitem
\bibitem[Stauber(2014)]{Stauber14}
Stauber,~T. Plasmonics in Dirac systems: from graphene to topological
  insulators. \emph{Journal of Physics: Condensed Matter} \textbf{2014},
  \emph{26}, 123201\relax
\mciteBstWouldAddEndPuncttrue
\mciteSetBstMidEndSepPunct{\mcitedefaultmidpunct}
{\mcitedefaultendpunct}{\mcitedefaultseppunct}\relax
\EndOfBibitem
\bibitem[Vald\'es~Aguilar \latin{et~al.}(2012)Vald\'es~Aguilar, Stier, Liu,
  Bilbro, George, Bansal, Wu, Cerne, Markelz, Oh, and Armitage]{Valdes12}
Vald\'es~Aguilar,~R.; Stier,~A.~V.; Liu,~W.; Bilbro,~L.~S.; George,~D.~K.;
  Bansal,~N.; Wu,~L.; Cerne,~J.; Markelz,~A.~G.; Oh,~S.; Armitage,~N.~P.
  Terahertz Response and Colossal Kerr Rotation from the Surface States of the
  Topological Insulator ${\mathrm{Bi}}_{2}{\mathrm{Se}}_{3}$. \emph{Phys. Rev.
  Lett.} \textbf{2012}, \emph{108}, 087403\relax
\mciteBstWouldAddEndPuncttrue
\mciteSetBstMidEndSepPunct{\mcitedefaultmidpunct}
{\mcitedefaultendpunct}{\mcitedefaultseppunct}\relax
\EndOfBibitem
\bibitem[Di~Pietro \latin{et~al.}(2013)Di~Pietro, Ortolani, Limaj, Di~Gaspare,
  Giliberti, Giorgianni, Brahlek, Bansal, Koirala, Oh, Calvani, and
  Lupi]{DiPietro13}
Di~Pietro,~P.; Ortolani,~M.; Limaj,~O.; Di~Gaspare,~A.; Giliberti,~V.;
  Giorgianni,~F.; Brahlek,~M.; Bansal,~N.; Koirala,~N.; Oh,~S.; Calvani,~P.;
  Lupi,~S. Observation of Dirac plasmons in a topological insulator.
  \emph{Nature Nanotechnol.} \textbf{2013}, \emph{8}, 556--560\relax
\mciteBstWouldAddEndPuncttrue
\mciteSetBstMidEndSepPunct{\mcitedefaultmidpunct}
{\mcitedefaultendpunct}{\mcitedefaultseppunct}\relax
\EndOfBibitem
\bibitem[Autore \latin{et~al.}(2015)Autore, D'Apuzzo, Di~Gaspare, Giliberti,
  Limaj, Roy, Brahlek, Koirala, Oh, Garc\'{\i}a~de Abajo, and Lupi]{Autore15a}
Autore,~M.; D'Apuzzo,~F.; Di~Gaspare,~A.; Giliberti,~V.; Limaj,~O.; Roy,~P.;
  Brahlek,~M.; Koirala,~N.; Oh,~S.; Garc\'{\i}a~de Abajo,~F.~J.; Lupi,~S.
  Plasmon–Phonon Interactions in Topological Insulator Microrings.
  \emph{Advanced Optical Materials} \textbf{2015}, \emph{3}, 1257--1263\relax
\mciteBstWouldAddEndPuncttrue
\mciteSetBstMidEndSepPunct{\mcitedefaultmidpunct}
{\mcitedefaultendpunct}{\mcitedefaultseppunct}\relax
\EndOfBibitem
\bibitem[Autore \latin{et~al.}(2015)Autore, Engelkamp, D’Apuzzo, Gaspare,
  Pietro, Vecchio, Brahlek, Koirala, Oh, and Lupi]{Autore15b}
Autore,~M.; Engelkamp,~H.; D’Apuzzo,~F.; Gaspare,~A.~D.; Pietro,~P.~D.;
  Vecchio,~I.~L.; Brahlek,~M.; Koirala,~N.; Oh,~S.; Lupi,~S. Observation of
  Magnetoplasmons in Bi2Se3 Topological Insulator. \emph{ACS Photonics}
  \textbf{2015}, \emph{2}, 1231--1235\relax
\mciteBstWouldAddEndPuncttrue
\mciteSetBstMidEndSepPunct{\mcitedefaultmidpunct}
{\mcitedefaultendpunct}{\mcitedefaultseppunct}\relax
\EndOfBibitem
\bibitem[Post \latin{et~al.}(2015)Post, Chapler, Liu, Wu, Stinson, Goldflam,
  Richardella, Lee, Reijnders, Burch, Fogler, Samarth, and Basov]{Post15}
Post,~K.~W.; Chapler,~B.~C.; Liu,~M.~K.; Wu,~J.~S.; Stinson,~H.~T.;
  Goldflam,~M.~D.; Richardella,~A.~R.; Lee,~J.~S.; Reijnders,~A.~A.;
  Burch,~K.~S.; Fogler,~M.~M.; Samarth,~N.; Basov,~D.~N. Sum-Rule Constraints
  on the Surface State Conductance of Topological Insulators. \emph{Phys. Rev.
  Lett.} \textbf{2015}, \emph{115}, 116804\relax
\mciteBstWouldAddEndPuncttrue
\mciteSetBstMidEndSepPunct{\mcitedefaultmidpunct}
{\mcitedefaultendpunct}{\mcitedefaultseppunct}\relax
\EndOfBibitem
\bibitem[Sim \latin{et~al.}(2015)Sim, Jang, Koirala, Brahlek, Moon, Sung, Park,
  Cha, Oh, Jo, Ahn, and Choi]{Sim15}
Sim,~S.; Jang,~H.; Koirala,~N.; Brahlek,~M.; Moon,~J.; Sung,~J.~H.; Park,~J.;
  Cha,~S.; Oh,~S.; Jo,~M.-H.; Ahn,~J.-H.; Choi,~H. Ultra-high modulation depth
  exceeding 2,400{\%} in optically controlled topological surface plasmons.
  \emph{Nat Commun} \textbf{2015}, \emph{6}, 8814\relax
\mciteBstWouldAddEndPuncttrue
\mciteSetBstMidEndSepPunct{\mcitedefaultmidpunct}
{\mcitedefaultendpunct}{\mcitedefaultseppunct}\relax
\EndOfBibitem
\bibitem[Zhao \latin{et~al.}(2015)Zhao, Bosman, Danesh, Zeng, Song, Darma,
  Rusydi, Lin, Qiu, and Loh]{Zhao15}
Zhao,~M.; Bosman,~M.; Danesh,~M.; Zeng,~M.; Song,~P.; Darma,~Y.; Rusydi,~A.;
  Lin,~H.; Qiu,~C.-W.; Loh,~K.~P. Visible Surface Plasmon Modes in Single
  Bi2Te3 Nanoplate. \emph{Nano Letters} \textbf{2015}, \emph{15}, 8331--8335,
  PMID: 26569579\relax
\mciteBstWouldAddEndPuncttrue
\mciteSetBstMidEndSepPunct{\mcitedefaultmidpunct}
{\mcitedefaultendpunct}{\mcitedefaultseppunct}\relax
\EndOfBibitem
\bibitem[Politano \latin{et~al.}(2015)Politano, Silkin, Nechaev, Vitiello,
  Viti, Aliev, Babanly, Chiarello, Echenique, and Chulkov]{Politano15}
Politano,~A.; Silkin,~V.~M.; Nechaev,~I.~A.; Vitiello,~M.~S.; Viti,~L.;
  Aliev,~Z.~S.; Babanly,~M.~B.; Chiarello,~G.; Echenique,~P.~M.; Chulkov,~E.~V.
  Interplay of Surface and Dirac Plasmons in Topological Insulators: The Case
  of ${\mathrm{Bi}}_{2}{\mathrm{Se}}_{3}$. \emph{Phys. Rev. Lett.}
  \textbf{2015}, \emph{115}, 216802\relax
\mciteBstWouldAddEndPuncttrue
\mciteSetBstMidEndSepPunct{\mcitedefaultmidpunct}
{\mcitedefaultendpunct}{\mcitedefaultseppunct}\relax
\EndOfBibitem
\bibitem[Wu \latin{et~al.}(2015)Wu, Tse, Brahlek, Morris, Aguilar, Koirala, Oh,
  and Armitage]{Wu15}
Wu,~L.; Tse,~W.-K.; Brahlek,~M.; Morris,~C.~M.; Aguilar,~R.~V.; Koirala,~N.;
  Oh,~S.; Armitage,~N.~P. High-Resolution Faraday Rotation and Electron-Phonon
  Coupling in Surface States of the Bulk-Insulating Topological Insulator
  ${\mathrm{Cu}}_{0.02}{\mathrm{Bi}}_{2}{\mathrm{Se}}_{3}$. \emph{Phys. Rev.
  Lett.} \textbf{2015}, \emph{115}, 217602\relax
\mciteBstWouldAddEndPuncttrue
\mciteSetBstMidEndSepPunct{\mcitedefaultmidpunct}
{\mcitedefaultendpunct}{\mcitedefaultseppunct}\relax
\EndOfBibitem
\bibitem[Autore \latin{et~al.}(2016)Autore, Giorgianni, D{'}Apuzzo, Di~Gaspare,
  Lo~Vecchio, Brahlek, Koirala, Oh, Schade, Ortolani, and Lupi]{Autore16}
Autore,~M.; Giorgianni,~F.; D{'}Apuzzo,~F.; Di~Gaspare,~A.; Lo~Vecchio,~I.;
  Brahlek,~M.; Koirala,~N.; Oh,~S.; Schade,~U.; Ortolani,~M.; Lupi,~S.
  Topologically protected Dirac plasmons and their evolution across the Quantum
  Phase Transition in (Bi1-xInx)2Se3 Topological Insulator. \emph{Nanoscale}
  \textbf{2016}, \emph{8}, 4667--4671\relax
\mciteBstWouldAddEndPuncttrue
\mciteSetBstMidEndSepPunct{\mcitedefaultmidpunct}
{\mcitedefaultendpunct}{\mcitedefaultseppunct}\relax
\EndOfBibitem
\bibitem[Viti \latin{et~al.}(2016)Viti, Coquillat, Politano, Kokh, Aliev,
  Babanly, Tereshchenko, Knap, Chulkov, and Vitiello]{Viti16}
Viti,~L.; Coquillat,~D.; Politano,~A.; Kokh,~K.~A.; Aliev,~Z.~S.;
  Babanly,~M.~B.; Tereshchenko,~O.~E.; Knap,~W.; Chulkov,~E.~V.;
  Vitiello,~M.~S. Plasma-Wave Terahertz Detection Mediated by Topological
  Insulators Surface States. \emph{Nano Letters} \textbf{2016}, \emph{16},
  80--87, PMID: 26678677\relax
\mciteBstWouldAddEndPuncttrue
\mciteSetBstMidEndSepPunct{\mcitedefaultmidpunct}
{\mcitedefaultendpunct}{\mcitedefaultseppunct}\relax
\EndOfBibitem
\bibitem[Autore \latin{et~al.}(2017)Autore, Pietro, Gaspare, D{'}Apuzzo,
  Giorgianni, Brahlek, Koirala, Oh, and Lupi]{Autore17}
Autore,~M.; Pietro,~P.~D.; Gaspare,~A.~D.; D{'}Apuzzo,~F.; Giorgianni,~F.;
  Brahlek,~M.; Koirala,~N.; Oh,~S.; Lupi,~S. Terahertz plasmonic excitations in
  Bi 2 Se 3 topological insulator. \emph{Journal of Physics: Condensed Matter}
  \textbf{2017}, \emph{29}, 183002\relax
\mciteBstWouldAddEndPuncttrue
\mciteSetBstMidEndSepPunct{\mcitedefaultmidpunct}
{\mcitedefaultendpunct}{\mcitedefaultseppunct}\relax
\EndOfBibitem
\bibitem[Politano \latin{et~al.}(2017)Politano, Viti, and Vitiello]{Politano17}
Politano,~A.; Viti,~L.; Vitiello,~M.~S. Optoelectronic devices, plasmonics, and
  photonics with topological insulators. \emph{APL Materials} \textbf{2017},
  \emph{5}, 035504\relax
\mciteBstWouldAddEndPuncttrue
\mciteSetBstMidEndSepPunct{\mcitedefaultmidpunct}
{\mcitedefaultendpunct}{\mcitedefaultseppunct}\relax
\EndOfBibitem
\bibitem[Juergens \latin{et~al.}(2014)Juergens, Michetti, and
  Trauzettel]{Juergens14a}
Juergens,~S.; Michetti,~P.; Trauzettel,~B. Plasmons due to the Interplay of
  Dirac and Schr\"odinger Fermions. \emph{Phys. Rev. Lett.} \textbf{2014},
  \emph{112}, 076804\relax
\mciteBstWouldAddEndPuncttrue
\mciteSetBstMidEndSepPunct{\mcitedefaultmidpunct}
{\mcitedefaultendpunct}{\mcitedefaultseppunct}\relax
\EndOfBibitem
\bibitem[Juergens \latin{et~al.}(2014)Juergens, Michetti, and
  Trauzettel]{Juergens14b}
Juergens,~S.; Michetti,~P.; Trauzettel,~B. Screening properties and plasmons of
  Hg(Cd)Te quantum wells. \emph{Phys. Rev. B} \textbf{2014}, \emph{90},
  115425\relax
\mciteBstWouldAddEndPuncttrue
\mciteSetBstMidEndSepPunct{\mcitedefaultmidpunct}
{\mcitedefaultendpunct}{\mcitedefaultseppunct}\relax
\EndOfBibitem
\bibitem[Qi \latin{et~al.}(2014)Qi, Liu, and Xie]{Qi15}
Qi,~J.; Liu,~H.; Xie,~X.~C. Surface plasmon polaritons in topological
  insulators. \emph{Phys. Rev. B} \textbf{2014}, \emph{89}, 155420\relax
\mciteBstWouldAddEndPuncttrue
\mciteSetBstMidEndSepPunct{\mcitedefaultmidpunct}
{\mcitedefaultendpunct}{\mcitedefaultseppunct}\relax
\EndOfBibitem
\bibitem[Wu \latin{et~al.}(2015)Wu, Basov, and Fogler]{Wu15b}
Wu,~J.-S.; Basov,~D.~N.; Fogler,~M.~M. Topological insulators are tunable
  waveguides for hyperbolic polaritons. \emph{Phys. Rev. B} \textbf{2015},
  \emph{92}, 205430\relax
\mciteBstWouldAddEndPuncttrue
\mciteSetBstMidEndSepPunct{\mcitedefaultmidpunct}
{\mcitedefaultendpunct}{\mcitedefaultseppunct}\relax
\EndOfBibitem
\bibitem[Lin \latin{et~al.}(2015)Lin, Lin, Chi, Wu, Cheng, Tseng, He, Wu, Lee,
  and Lin]{Lin15}
Lin,~Y.-H.; Lin,~S.-F.; Chi,~Y.-C.; Wu,~C.-L.; Cheng,~C.-H.; Tseng,~W.-H.;
  He,~J.-H.; Wu,~C.-I.; Lee,~C.-K.; Lin,~G.-R. Using n- and p-Type Bi2Te3
  Topological Insulator Nanoparticles To Enable Controlled Femtosecond
  Mode-Locking of Fiber Lasers. \emph{ACS Photonics} \textbf{2015}, \emph{2},
  481--490\relax
\mciteBstWouldAddEndPuncttrue
\mciteSetBstMidEndSepPunct{\mcitedefaultmidpunct}
{\mcitedefaultendpunct}{\mcitedefaultseppunct}\relax
\EndOfBibitem
\bibitem[Siroki \latin{et~al.}(2016)Siroki, Lee, Haynes, and
  Giannini]{Siroki17}
Siroki,~G.; Lee,~D. K.~K.; Haynes,~P.~D.; Giannini,~V. Single-electron induced
  surface plasmons on a topological nanoparticle. \textbf{2016}, \emph{7},
  12375 EP --\relax
\mciteBstWouldAddEndPuncttrue
\mciteSetBstMidEndSepPunct{\mcitedefaultmidpunct}
{\mcitedefaultendpunct}{\mcitedefaultseppunct}\relax
\EndOfBibitem
\bibitem[Lee(2009)]{Lee09}
Lee,~D.-H. Surface States of Topological Insulators: The Dirac Fermion in
  Curved Two-Dimensional Spaces. \emph{Phys. Rev. Lett.} \textbf{2009},
  \emph{103}, 196804\relax
\mciteBstWouldAddEndPuncttrue
\mciteSetBstMidEndSepPunct{\mcitedefaultmidpunct}
{\mcitedefaultendpunct}{\mcitedefaultseppunct}\relax
\EndOfBibitem
\bibitem[Silvestrov \latin{et~al.}(2012)Silvestrov, Brouwer, and
  Mishchenko]{Silvestrov12}
Silvestrov,~P.~G.; Brouwer,~P.~W.; Mishchenko,~E.~G. Spin and charge structure
  of the surface states in topological insulators. \emph{Phys. Rev. B}
  \textbf{2012}, \emph{86}, 075302\relax
\mciteBstWouldAddEndPuncttrue
\mciteSetBstMidEndSepPunct{\mcitedefaultmidpunct}
{\mcitedefaultendpunct}{\mcitedefaultseppunct}\relax
\EndOfBibitem
\bibitem[Wunsch \latin{et~al.}(2006)Wunsch, Stauber, Sols, and
  Guinea]{Wunsch06}
Wunsch,~B.; Stauber,~T.; Sols,~F.; Guinea,~F. Dynamical polarization of
  graphene at finite doping. \emph{New Journal of Physics} \textbf{2006},
  \emph{8}, 318\relax
\mciteBstWouldAddEndPuncttrue
\mciteSetBstMidEndSepPunct{\mcitedefaultmidpunct}
{\mcitedefaultendpunct}{\mcitedefaultseppunct}\relax
\EndOfBibitem
\bibitem[Hwang and Das~Sarma(2007)Hwang, and Das~Sarma]{Hwang07}
Hwang,~E.~H.; Das~Sarma,~S. Dielectric function, screening, and plasmons in
  two-dimensional graphene. \emph{Phys. Rev. B} \textbf{2007}, \emph{75},
  205418\relax
\mciteBstWouldAddEndPuncttrue
\mciteSetBstMidEndSepPunct{\mcitedefaultmidpunct}
{\mcitedefaultendpunct}{\mcitedefaultseppunct}\relax
\EndOfBibitem
\bibitem[Profumo \latin{et~al.}(2012)Profumo, Asgari, Polini, and
  MacDonald]{Profumo12}
Profumo,~R. E.~V.; Asgari,~R.; Polini,~M.; MacDonald,~A.~H. Double-layer
  graphene and topological insulator thin-film plasmons. \emph{Phys. Rev. B}
  \textbf{2012}, \emph{85}, 085443\relax
\mciteBstWouldAddEndPuncttrue
\mciteSetBstMidEndSepPunct{\mcitedefaultmidpunct}
{\mcitedefaultendpunct}{\mcitedefaultseppunct}\relax
\EndOfBibitem
\bibitem[Badalyan and Peeters(2012)Badalyan, and Peeters]{Badalyan12}
Badalyan,~S.~M.; Peeters,~F.~M. Effect of nonhomogenous dielectric background
  on the plasmon modes in graphene double-layer structures at finite
  temperatures. \emph{Phys. Rev. B} \textbf{2012}, \emph{85}, 195444\relax
\mciteBstWouldAddEndPuncttrue
\mciteSetBstMidEndSepPunct{\mcitedefaultmidpunct}
{\mcitedefaultendpunct}{\mcitedefaultseppunct}\relax
\EndOfBibitem
\bibitem[Scharf and Matos-Abiague(2012)Scharf, and Matos-Abiague]{Scharf12}
Scharf,~B.; Matos-Abiague,~A. Coulomb drag between massless and massive
  fermions. \emph{Phys. Rev. B} \textbf{2012}, \emph{86}, 115425\relax
\mciteBstWouldAddEndPuncttrue
\mciteSetBstMidEndSepPunct{\mcitedefaultmidpunct}
{\mcitedefaultendpunct}{\mcitedefaultseppunct}\relax
\EndOfBibitem
\bibitem[Stauber and G\'omez-Santos(2012)Stauber, and
  G\'omez-Santos]{Stauber12}
Stauber,~T.; G\'omez-Santos,~G. Plasmons in layered structures including
  graphene. \emph{New Journal of Physics} \textbf{2012}, \emph{14},
  105018\relax
\mciteBstWouldAddEndPuncttrue
\mciteSetBstMidEndSepPunct{\mcitedefaultmidpunct}
{\mcitedefaultendpunct}{\mcitedefaultseppunct}\relax
\EndOfBibitem
\bibitem[Santoro and Giuliani(1988)Santoro, and Giuliani]{Santoro88}
Santoro,~G.~E.; Giuliani,~G.~F. Acoustic plasmons in a conducting double layer.
  \emph{Phys. Rev. B} \textbf{1988}, \emph{37}, 937--940\relax
\mciteBstWouldAddEndPuncttrue
\mciteSetBstMidEndSepPunct{\mcitedefaultmidpunct}
{\mcitedefaultendpunct}{\mcitedefaultseppunct}\relax
\EndOfBibitem
\bibitem[Ameen~Poyli \latin{et~al.}()Ameen~Poyli, Hrton, Nechaev, Nikitin,
  Echenique, Silkin, Aizpurua, and Esteban]{Ameen17}
Ameen~Poyli,~M.; Hrton,~M.; Nechaev,~M. I.~A.; Nikitin,~A.; Echenique,~P.~M.;
  Silkin,~V.~M.; Aizpurua,~J.; Esteban,~R. Controlling surface charge and spin
  density oscillations by Dirac plasmon interaction in thin topological
  insulators. \emph{arXiv:1707.03050} \relax
\mciteBstWouldAddEndPunctfalse
\mciteSetBstMidEndSepPunct{\mcitedefaultmidpunct}
{}{\mcitedefaultseppunct}\relax
\EndOfBibitem
\bibitem[Bondarev and Shalaev()Bondarev, and Shalaev]{Bondarev17}
Bondarev,~I.~V.; Shalaev,~V.~M. Universal features of the optical properties of
  ultrathin plasmonic films. \emph{arXiv:1708.03553} \relax
\mciteBstWouldAddEndPunctfalse
\mciteSetBstMidEndSepPunct{\mcitedefaultmidpunct}
{}{\mcitedefaultseppunct}\relax
\EndOfBibitem
\bibitem[Bansal \latin{et~al.}(2012)Bansal, Kim, Brahlek, Edrey, and
  Oh]{Bansal12}
Bansal,~N.; Kim,~Y.~S.; Brahlek,~M.; Edrey,~E.; Oh,~S. Thickness-Independent
  Transport Channels in Topological Insulator
  ${\mathrm{Bi}}_{2}{\mathrm{Se}}_{3}$ Thin Films. \emph{Phys. Rev. Lett.}
  \textbf{2012}, \emph{109}, 116804\relax
\mciteBstWouldAddEndPuncttrue
\mciteSetBstMidEndSepPunct{\mcitedefaultmidpunct}
{\mcitedefaultendpunct}{\mcitedefaultseppunct}\relax
\EndOfBibitem
\bibitem[Brey and Fertig(2007)Brey, and Fertig]{Brey07}
Brey,~L.; Fertig,~H.~A. Elementary electronic excitations in graphene
  nanoribbons. \emph{Phys. Rev. B} \textbf{2007}, \emph{75}, 125434\relax
\mciteBstWouldAddEndPuncttrue
\mciteSetBstMidEndSepPunct{\mcitedefaultmidpunct}
{\mcitedefaultendpunct}{\mcitedefaultseppunct}\relax
\EndOfBibitem
\bibitem[Egger \latin{et~al.}(2010)Egger, Zazunov, and Yeyati]{Egger10}
Egger,~R.; Zazunov,~A.; Yeyati,~A.~L. Helical Luttinger Liquid in Topological
  Insulator Nanowires. \emph{Phys. Rev. Lett.} \textbf{2010}, \emph{105},
  136403\relax
\mciteBstWouldAddEndPuncttrue
\mciteSetBstMidEndSepPunct{\mcitedefaultmidpunct}
{\mcitedefaultendpunct}{\mcitedefaultseppunct}\relax
\EndOfBibitem
\bibitem[Moroz \latin{et~al.}(2000)Moroz, Samokhin, and Barnes]{Moroz00}
Moroz,~A.~V.; Samokhin,~K.~V.; Barnes,~C. H.~W. Spin-Orbit Coupling in
  Interacting Quasi-One-Dimensional Electron Systems. \emph{Phys. Rev. Lett.}
  \textbf{2000}, \emph{84}, 4164--4167\relax
\mciteBstWouldAddEndPuncttrue
\mciteSetBstMidEndSepPunct{\mcitedefaultmidpunct}
{\mcitedefaultendpunct}{\mcitedefaultseppunct}\relax
\EndOfBibitem
\bibitem[Ju \latin{et~al.}(2011)Ju, Geng, Horng, Girit, Martin, Hao, Bechtel,
  Liang, Zettl, Shen, and Wang]{Ju12}
Ju,~L.; Geng,~B.; Horng,~J.; Girit,~C.; Martin,~M.; Hao,~Z.; Bechtel,~H.~A.;
  Liang,~X.; Zettl,~A.; Shen,~Y.~R.; Wang,~F. Graphene plasmonics for tunable
  terahertz metamaterials. \emph{Nature Nanotech.} \textbf{2011}, \emph{6},
  630\relax
\mciteBstWouldAddEndPuncttrue
\mciteSetBstMidEndSepPunct{\mcitedefaultmidpunct}
{\mcitedefaultendpunct}{\mcitedefaultseppunct}\relax
\EndOfBibitem
\bibitem[Yan \latin{et~al.}(2013)Yan, Low, Zhu, Wu, Freitag, Li, Guinea,
  Avouris, and Xia]{Yan13}
Yan,~H.; Low,~T.; Zhu,~W.; Wu,~Y.; Freitag,~M.; Li,~X.; Guinea,~F.;
  Avouris,~P.; Xia,~F. Damping pathways of mid-infrared plasmons in graphene
  nanostructures. \emph{Nat Photon} \textbf{2013}, \emph{7}, 394--399\relax
\mciteBstWouldAddEndPuncttrue
\mciteSetBstMidEndSepPunct{\mcitedefaultmidpunct}
{\mcitedefaultendpunct}{\mcitedefaultseppunct}\relax
\EndOfBibitem
\bibitem[Stordeur \latin{et~al.}(1992)Stordeur, Ketavong, Priemuth, Sobotta,
  and Riede]{Stordeur92}
Stordeur,~M.; Ketavong,~K.~K.; Priemuth,~A.; Sobotta,~H.; Riede,~V. Optical and
  Electrical Investigations of n-Type Bi2Se3 Single Crystals. \emph{physica
  status solidi (b)} \textbf{1992}, \emph{169}, 505--514\relax
\mciteBstWouldAddEndPuncttrue
\mciteSetBstMidEndSepPunct{\mcitedefaultmidpunct}
{\mcitedefaultendpunct}{\mcitedefaultseppunct}\relax
\EndOfBibitem
\bibitem[Nechaev \latin{et~al.}(2013)Nechaev, Hatch, Bianchi, Guan, Friedrich,
  Aguilera, Mi, Iversen, Bl\"ugel, Hofmann, and Chulkov]{Nechaev13}
Nechaev,~I.~A.; Hatch,~R.~C.; Bianchi,~M.; Guan,~D.; Friedrich,~C.;
  Aguilera,~I.; Mi,~J.~L.; Iversen,~B.~B.; Bl\"ugel,~S.; Hofmann,~P.;
  Chulkov,~E.~V. Evidence for a direct band gap in the topological insulator
  Bi${}_{2}$Se${}_{3}$ from theory and experiment. \emph{Phys. Rev. B}
  \textbf{2013}, \emph{87}, 121111\relax
\mciteBstWouldAddEndPuncttrue
\mciteSetBstMidEndSepPunct{\mcitedefaultmidpunct}
{\mcitedefaultendpunct}{\mcitedefaultseppunct}\relax
\EndOfBibitem
\bibitem[Bianchi \latin{et~al.}(2010)Bianchi, Guan, Bao, Mi, Iversen, King, and
  Hofmann]{Bianchi10}
Bianchi,~M.; Guan,~D.; Bao,~S.; Mi,~J.; Iversen,~B.~B.; King,~P.~D.;
  Hofmann,~P. Coexistence of the topological state and a two-dimensional
  electron gas on the surface of Bi2Se3. \emph{Nature Commun.} \textbf{2010},
  \emph{1}, 128\relax
\mciteBstWouldAddEndPuncttrue
\mciteSetBstMidEndSepPunct{\mcitedefaultmidpunct}
{\mcitedefaultendpunct}{\mcitedefaultseppunct}\relax
\EndOfBibitem
\bibitem[Stern(1967)]{Stern67}
Stern,~F. Polarizability of a Two-Dimensional Electron Gas. \emph{Phys. Rev.
  Lett.} \textbf{1967}, \emph{18}, 546--548\relax
\mciteBstWouldAddEndPuncttrue
\mciteSetBstMidEndSepPunct{\mcitedefaultmidpunct}
{\mcitedefaultendpunct}{\mcitedefaultseppunct}\relax
\EndOfBibitem
\bibitem[Nikitin \latin{et~al.}(2012)Nikitin, Guinea, Garcia-Vidal, and
  Martin-Moreno]{Nikitin12}
Nikitin,~A.~Y.; Guinea,~F.; Garcia-Vidal,~F.~J.; Martin-Moreno,~L. Surface
  plasmon enhanced absorption and suppressed transmission in periodic arrays of
  graphene ribbons. \emph{Phys. Rev. B} \textbf{2012}, \emph{85}, 081405\relax
\mciteBstWouldAddEndPuncttrue
\mciteSetBstMidEndSepPunct{\mcitedefaultmidpunct}
{\mcitedefaultendpunct}{\mcitedefaultseppunct}\relax
\EndOfBibitem
\bibitem[Christensen \latin{et~al.}(2012)Christensen, Manjavacas,
  Thongrattanasiri, Koppens, and Garc{\'\i}a~de Abajo]{Christensen12}
Christensen,~J.; Manjavacas,~A.; Thongrattanasiri,~S.; Koppens,~F. H.~L.;
  Garc{\'\i}a~de Abajo,~F.~J. Graphene Plasmon Waveguiding and Hybridization in
  Individual and Paired Nanoribbons. \emph{ACS Nano} \textbf{2012}, \emph{6},
  431--440\relax
\mciteBstWouldAddEndPuncttrue
\mciteSetBstMidEndSepPunct{\mcitedefaultmidpunct}
{\mcitedefaultendpunct}{\mcitedefaultseppunct}\relax
\EndOfBibitem
\bibitem[Strait \latin{et~al.}(2013)Strait, Nene, Chan, Manolatou, Tiwari,
  Rana, Kevek, and McEuen]{Strait13}
Strait,~J.~H.; Nene,~P.; Chan,~W.-M.; Manolatou,~C.; Tiwari,~S.; Rana,~F.;
  Kevek,~J.~W.; McEuen,~P.~L. Confined plasmons in graphene microstructures:
  Experiments and theory. \emph{Phys. Rev. B} \textbf{2013}, \emph{87},
  241410\relax
\mciteBstWouldAddEndPuncttrue
\mciteSetBstMidEndSepPunct{\mcitedefaultmidpunct}
{\mcitedefaultendpunct}{\mcitedefaultseppunct}\relax
\EndOfBibitem
\bibitem[Kim \latin{et~al.}(2009)Kim, Nah, Jo, Shahrjerdi, Colombo, Yao, Tutuc,
  and Banerjee]{Kim09}
Kim,~S.; Nah,~J.; Jo,~I.; Shahrjerdi,~D.; Colombo,~L.; Yao,~Z.; Tutuc,~E.;
  Banerjee,~S.~K. Realization of a high mobility dual-gated graphene
  field-effect transistor with Al[sub 2]O[sub 3] dielectric. \emph{Applied
  Physics Letters} \textbf{2009}, \emph{94}, 062107\relax
\mciteBstWouldAddEndPuncttrue
\mciteSetBstMidEndSepPunct{\mcitedefaultmidpunct}
{\mcitedefaultendpunct}{\mcitedefaultseppunct}\relax
\EndOfBibitem
\bibitem[Deshko \latin{et~al.}(2016)Deshko, Krusin-Elbaum, Menon, Khanikaev,
  and Trevino]{Deshko16}
Deshko,~Y.; Krusin-Elbaum,~L.; Menon,~V.; Khanikaev,~A.; Trevino,~J. Surface
  plasmon polaritons in topological insulator nano-films and superlattices.
  \emph{Opt. Express} \textbf{2016}, \emph{24}, 7398--7410\relax
\mciteBstWouldAddEndPuncttrue
\mciteSetBstMidEndSepPunct{\mcitedefaultmidpunct}
{\mcitedefaultendpunct}{\mcitedefaultseppunct}\relax
\EndOfBibitem
\bibitem[Abedinpour \latin{et~al.}(2011)Abedinpour, Vignale, Principi, Polini,
  Tse, and MacDonald]{Abedinpour11}
Abedinpour,~S.~H.; Vignale,~G.; Principi,~A.; Polini,~M.; Tse,~W.-K.;
  MacDonald,~A.~H. Drude weight, plasmon dispersion, and ac conductivity in
  doped graphene sheets. \emph{Phys. Rev. B} \textbf{2011}, \emph{84},
  045429\relax
\mciteBstWouldAddEndPuncttrue
\mciteSetBstMidEndSepPunct{\mcitedefaultmidpunct}
{\mcitedefaultendpunct}{\mcitedefaultseppunct}\relax
\EndOfBibitem
\bibitem[Levitov \latin{et~al.}(2013)Levitov, Shtyk, and Feigelman]{Levitov13}
Levitov,~L.~S.; Shtyk,~A.~V.; Feigelman,~M.~V. Electron-electron interactions
  and plasmon dispersion in graphene. \emph{Phys. Rev. B} \textbf{2013},
  \emph{88}, 235403\relax
\mciteBstWouldAddEndPuncttrue
\mciteSetBstMidEndSepPunct{\mcitedefaultmidpunct}
{\mcitedefaultendpunct}{\mcitedefaultseppunct}\relax
\EndOfBibitem
\bibitem[Stauber \latin{et~al.}(2017)Stauber, Parida, Trushin, Ulybyshev,
  Boyda, and Schliemann]{Stauber17}
Stauber,~T.; Parida,~P.; Trushin,~M.; Ulybyshev,~M.~V.; Boyda,~D.~L.;
  Schliemann,~J. Interacting Electrons in Graphene: Fermi Velocity
  Renormalization and Optical Response. \emph{Phys. Rev. Lett.} \textbf{2017},
  \emph{118}, 266801\relax
\mciteBstWouldAddEndPuncttrue
\mciteSetBstMidEndSepPunct{\mcitedefaultmidpunct}
{\mcitedefaultendpunct}{\mcitedefaultseppunct}\relax
\EndOfBibitem
\bibitem[Stauber \latin{et~al.}(2013)Stauber, San-Jose, and Brey]{StauberNJP13}
Stauber,~T.; San-Jose,~P.; Brey,~L. Optical conductivity, Drude weight and
  plasmons in twisted graphene bilayers. \emph{New Journal of Physics}
  \textbf{2013}, \emph{15}, 113050\relax
\mciteBstWouldAddEndPuncttrue
\mciteSetBstMidEndSepPunct{\mcitedefaultmidpunct}
{\mcitedefaultendpunct}{\mcitedefaultseppunct}\relax
\EndOfBibitem
\bibitem[Stauber and Kohler(2016)Stauber, and Kohler]{Stauber16}
Stauber,~T.; Kohler,~H. Quasi-Flat Plasmonic Bands in Twisted Bilayer Graphene.
  \emph{Nano Letters} \textbf{2016}, \emph{16}, 6844--6849\relax
\mciteBstWouldAddEndPuncttrue
\mciteSetBstMidEndSepPunct{\mcitedefaultmidpunct}
{\mcitedefaultendpunct}{\mcitedefaultseppunct}\relax
\EndOfBibitem
\bibitem[Roberts and Coon(1962)Roberts, and Coon]{Roberts62}
Roberts,~S.; Coon,~D.~D. Far-Infrared Properties of Quartz and Sapphire.
  \emph{J. Opt. Soc. Am.} \textbf{1962}, \emph{52}, 1023--1029\relax
\mciteBstWouldAddEndPuncttrue
\mciteSetBstMidEndSepPunct{\mcitedefaultmidpunct}
{\mcitedefaultendpunct}{\mcitedefaultseppunct}\relax
\EndOfBibitem
\bibitem[Scholz \latin{et~al.}(2012)Scholz, Stauber, and Schliemann]{Scholz12}
Scholz,~A.; Stauber,~T.; Schliemann,~J. Dielectric function, screening, and
  plasmons of graphene in the presence of spin-orbit interactions. \emph{Phys.
  Rev. B} \textbf{2012}, \emph{86}, 195424\relax
\mciteBstWouldAddEndPuncttrue
\mciteSetBstMidEndSepPunct{\mcitedefaultmidpunct}
{\mcitedefaultendpunct}{\mcitedefaultseppunct}\relax
\EndOfBibitem
\bibitem[Jia \latin{et~al.}()Jia, Zhang, Sankar, Chou, Wang, Kempa, Plummer,
  Zhang, Zhu, and Guo]{Jia17}
Jia,~X.; Zhang,~S.; Sankar,~R.; Chou,~F.-C.; Wang,~W.; Kempa,~K.;
  Plummer,~E.~W.; Zhang,~J.; Zhu,~X.; Guo,~J. An Anomalous Acoustic Plasmon
  Mode from Topologically Protected States. \emph{arXiv:1705.01685} \relax
\mciteBstWouldAddEndPunctfalse
\mciteSetBstMidEndSepPunct{\mcitedefaultmidpunct}
{}{\mcitedefaultseppunct}\relax
\EndOfBibitem
\bibitem[Scalapino \latin{et~al.}(1993)Scalapino, White, and
  Zhang]{Scalapino93}
Scalapino,~D.~J.; White,~S.~R.; Zhang,~S. Insulator, metal, or superconductor:
  The criteria. \emph{Phys. Rev. B} \textbf{1993}, \emph{47}, 7995--8007\relax
\mciteBstWouldAddEndPuncttrue
\mciteSetBstMidEndSepPunct{\mcitedefaultmidpunct}
{\mcitedefaultendpunct}{\mcitedefaultseppunct}\relax
\EndOfBibitem
\bibitem[Stauber and G\'omez-Santos(2010)Stauber, and
  G\'omez-Santos]{Stauber10b}
Stauber,~T.; G\'omez-Santos,~G. Dynamical current-current correlation of the
  hexagonal lattice and graphene. \emph{Phys. Rev. B} \textbf{2010}, \emph{82},
  155412\relax
\mciteBstWouldAddEndPuncttrue
\mciteSetBstMidEndSepPunct{\mcitedefaultmidpunct}
{\mcitedefaultendpunct}{\mcitedefaultseppunct}\relax
\EndOfBibitem
\end{mcitethebibliography}
\end{document}